\documentclass[iop,preprint,a4paper,amsmath,amssymb,floatfix]{revtex4}
\usepackage{amssymb}
\usepackage{dsfont}
\usepackage{bm}

\usepackage[T1]{fontenc}
\usepackage{pdfpages}

\usepackage[breaklinks=true,colorlinks=true,linkcolor=blue,urlcolor=blue,citecolor=blue]{hyperref}
\usepackage{color}
\newcommand{\beq}{\begin{equation}}
\newcommand{\eeq}{\end{equation}}
\newcommand{\beqar}{\begin{eqnarray}}
\newcommand{\eeqar}{\end{eqnarray}}
\newcommand{\bea}{\begin{eqnarray}}
\newcommand{\eea}{\end{eqnarray}}
\newcommand{\bcen}{\begin{center}}
\newcommand{\ecen}{\end{center}}

\newcommand{\bra}[1]{\left< #1 \right|}
\newcommand{\ket}[1]{\left| #1 \right>}

\newcommand{\f}[2]{\frac{#1}{#2}}

\newcommand{\pd}[2]{\frac {\partial #1}{\partial #2}}
\newcommand{\ppd}[2]{\frac {\partial^2 #1}{\partial^2 #2}}
\newcommand{\brac}[1]{(#1)}
\newcommand{\sbrac}[1]{[#1]}
\newcommand{\mean}[1]{\langle {#1} \rangle}
\newcommand{\infint}{\int_{-\infty}^{\infty}}
\newcommand{\ra}{\rightarrow}

\begin{document}
\title[Stochastic laser cooling enabled by many body effects]{Stochastic laser cooling enabled by many body effects}

\author{Roie Dann and Ronnie Kosloff}
\address{The Institute of Chemistry, The Hebrew University of Jerusalem, Jerusalem 91904, Israel}
\email{roie.dann@mail.huji.ac.il}

\begin{abstract}
A novel  laser cooling mechanism based on many body effects is presented. 
The method can be applicable for cooling a large class of atoms and molecules
in higher density than commonly excepted by existing methods.
The cooling mechanism relies on the collective encounters of particle and light. 
Stochastic events between the  particles and photons as well as a collective effect give rise to  energy transfer between these media.  
Such mechanism relies on multiple light-matter encounters, therefore requiring a sufficient particle density, $\rho \sim 10^{14} \text{cm}^{-3}$. 
This is an advantage for experiments where high phase space density is required.
A second tuning laser can be added increasing the applicability to many types of atoms and molecules. 
This tuning laser changes the inter-particle potential by inducing an AC stark effect. 
As a result the required trapping density can be reduced down to $\rho \sim 10^6 \text{cm}^{-3}$.
Simulations of phase space distributions were performed comparing different particle densities, trap potentials and light field intensity profiles. 
The modelling shows efficient cooling rates up to $~10^{2} \text{K/s}$  for a dense ensemble of $~^{87}$Rb atoms, and cooling rates up to $~6\cdot 10^{2} \text{K/s}$ when adding an additional tuning source.   
\end{abstract}

\maketitle

\section{Introduction}
\label{sec:intro}
Atom-photon interactions has been a major research topic in physics. Its origins can be traced to Kepler who as early as 1619, suggested that light may have a mechanical effect, when observing that a comet's tail is always pointing away from the sun \cite{kepler1619cometis}. Later on Maxwell suggested a phenomena known as "light pressure", 
pressure exerted on a surface when exposed to electro-magnetic radiation \cite{maxwell1862iii}. The topic was revolutionized by two papers (1909 and 1916), when Einstein, following Planck's law of black body radiation, showed that light energy quanta must carry a momentum set by $p = \frac{h}{\lambda}$  \cite{einstein1909development,einstein1917mitteilungen, einstein1917quantentheorie}.

Many experimental realizations exploiting the photon momentum have been performed \cite{metcalf2012laser}. 
In the present study we exploit this momentum transfer for cooling an ensemble of colliding atoms. 
The frequency shift during the collision is the source of momentum exchange between the particles and light.

The first suggestion for cooling atoms via photon-atom interactions was proposed by H.E.D. Scovil in 1959 \cite{scovil1959three}. Scovil pioneered a quantum thermodynamic approach to laser cooling thus introducing the first quantum refrigerator. Further advancement in laser cooling was not recorded until more than a decade later. In 1975, simultaneously and independently of Scovil's work, two groups of Wineland and Dehmelt, \cite{wineland1975proposed} as well as Hansch and Schawlow \cite{hansch1975cooling} introduced new theories for laser cooling. Wineland's and Dehmelt's work treat the cooling of ions in an ion trap, and Hansch's and Schawlow's theory concentrates on neutral atoms. The initial theory proposed by Wineland and Dehmelt, known as the Doppler Cooling, involves energy transfer from the atomic media to photons depending on the relative velocity of the atoms to the light source. 
The Doppler Cooling theory predicts a minimum temperature known as the 'Doppler limit' which for Sodium and Rubidium atoms amounts to $240 \mu \text{K}$ and $146 \mu \text{K}$, respectively  \cite{steck2001rubidium,steck2000sodium}. When experimental studies based on the theory took place the cooling was unexpectedly efficient and led to temperatures below the Doppler limit, raising theoretical questions concerning the underlying mechanism. 

A significant effort was aimed at extending or replacing the Doppler cooling theory. Diverse theories were developed such as Raman cooling, cavity mediated cooling, and Sisyphus cooling \cite{kasevich1992laser, ketterle1992slowing,diedrich1989laser, kerman2000beyond, dalibard1985dressed, dalibard1989laser, neuhauser1978optical,leibfried2003quantum,boozer2006cooling,metcalf2007laser,breeden1981stark}. All proposed theories describe different mechanisms of energy transfer between atomic and photon media, and take advantage of the electro-magnetic waves' global character, enabling  transport of energy away from the atomic medium.

The Sisyphus cooling theory was proposed in 1989 by Cohen Tannoudji and Jean Dalibard\cite{dalibard1985dressed}. The theory involves two interfering laser beams creating a standing wave with a polarization gradient oscillating spatially between three polarizations $\sigma^{+}$, $\pi$ and $\sigma^{-}$. The periodic potential imposed on the atoms affects the ground and excited states differently, resulting in a varying energy gap alternating spatially. Such spatial dependence of the energy gap allows an average energy transfer from the moving atoms via repetitive excitations. In conjunction with the theoretical work, experimental research achieved nano Kelvin temperatures, setting the stage for  the materialization of Bose-Einstein condensates by additional evaporative cooling \cite{anderson1995observation, davis1995bose,zwierlein2003observation}.

In the present  cooling theory  we change the focus from a single atom picture to a collective many body approach in a dense particle medium.
A new cooling mechanism is proposed converting kinetic to optical energy,  based on a stochastic modulation of the emission frequency 
due to relative motion of the particles.

The present  cooling theory  is based on  high particle density when collisions occur. It differs from the well established  laser cooling mechanism  appropriate for a sparse  medium. 
The high density is advantageous for experiments where a large phase space density is desired.
It has been claimed \cite{gallagher1989exoergic} that inter-atomic collisions in high densities will lead to trap loss or heating.
Nevertheless, collective effects of light trapped in the particle medium have been reported in ultracold medium 
\cite{walker1990collective,ketterle1993high,townsend1995phase},  experimentally demonstrating that dense cold samples are feasible. 
The trapped light generates an internal pressure which leads to the expansion of the atomic cloud \cite{pruvost2000expansion,bachelard2016collective,guerin2017light}.
More recent experiments \cite{camara2014scaling}  have observed these collective effects in very large magnetic-optical-traps.
In cooling, the high density effects should become our ally and not our foe.

Theoretical approaches to model these phenomena are based on continuum hydrodynamical theories \cite{pruvost2000expansion,rodrigues2016collective}. We will adopt such a hydrodynamic description for our cooling theory in addition to absorption and emission properties which depend on the atomic properties and density.

The theory is demonstrated on cooling Rubidium 87 atoms. Rb has been the workhorse of ultra-cold atomic physics, due to 
favourable properties such as convenient vapour pressure at room temperature and a large absorption cross section and a large scattering length for S-wave scattering.
The cooling scheme can also be applied to a variety  of atomic and molecular systems 
where an approximate closed cycle transition can be found and the particle density is dominated by two-body collisions.
To expand the possible cooling candidates  we suggest adding an additional tuning laser which modifies the collision parameters.
This laser also enables to reduce the  inter-particle density.  Sympathetic cooling of a mixture of species can also become applicable.

The Stochastic Cooling theory is introduced at the beginning of section \ref{sec:Theory}
followed by an explanation of the modelling technique, and derivation of the different model variables in Sections \ref{sec:Modeling}. Section \ref{sec:Results A} presents the results of the Stochastic laser cooling theory. An additional generalized scheme, Enhanced  Stochastic Cooling theory, which allows efficient cooling of non alkali atoms and molecules is presented in Section \ref{sec:Enhanced}. Following the theoretical method is a discussion and conclusions.

\section{Stochastic Cooling theory}
\label{sec:Theory}
\subsection{Laser cooling scheme}

\begin{figure}[htb!]
\centering
 \includegraphics[scale=0.4]{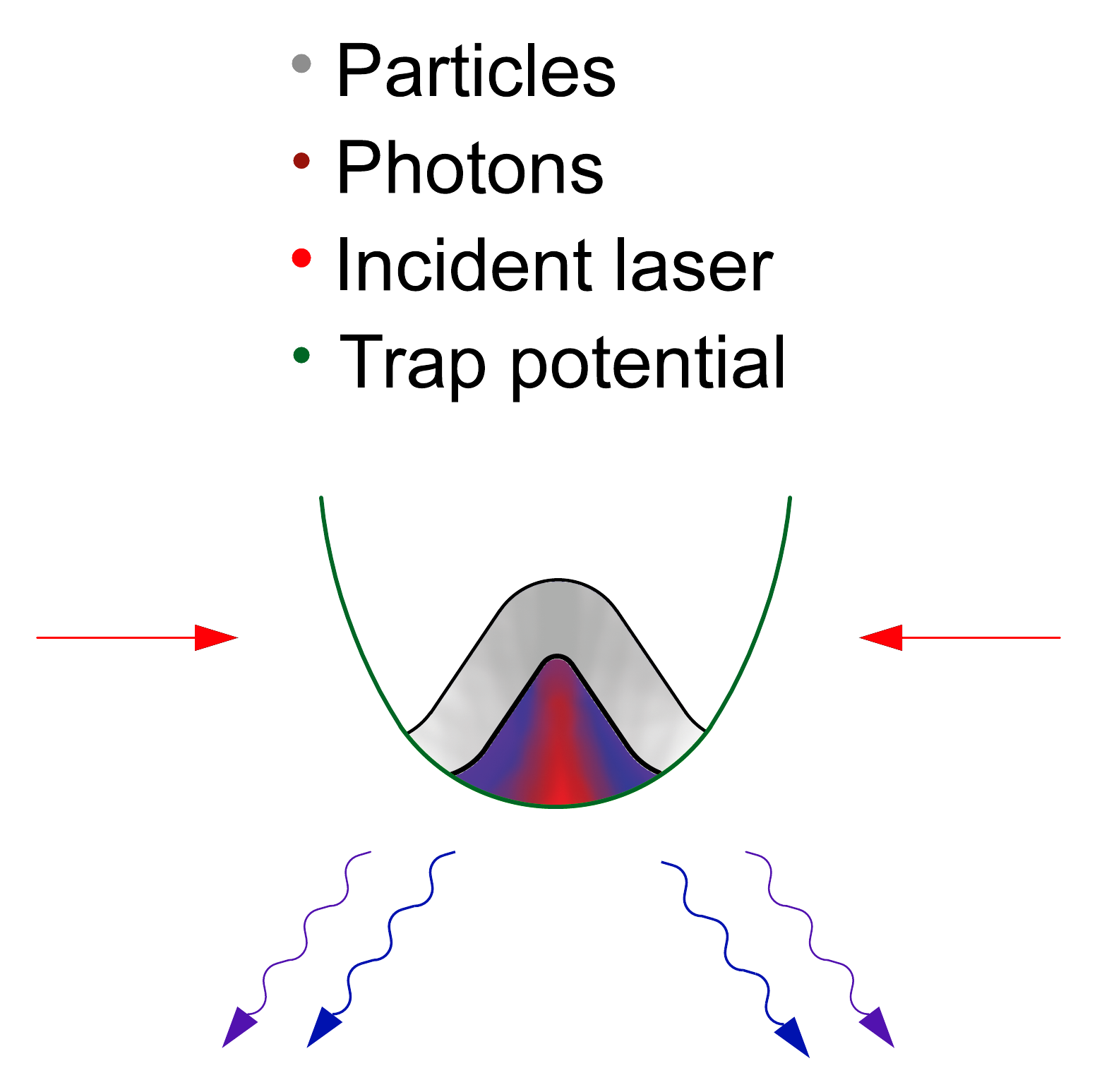} 

\caption{\label{figureone}$~^{87}\mbox{Rb}$  atoms are confined by a MOT (green), 
a laser source, detuned slightly from resonance (red), is applied to the particle ensemble (represented in grey). 
The photons are confined to the trap due to repeated absorption. They diffuse through the atomic media until they escape the trap blue detuned 
with respect to the incident laser (shown in blue and purple arrows). 
Inter-particle collisions modify the absorption cross section and locally equilibrate the kinetic energy.}
 \label{cooling_scheme}
\end{figure}

The prerequisite of the scheme is a trapping potential able to confine and isolate a dense ensemble of gas phase atoms or molecules.
The cooling is based on applying beams of light to the interior of the trap detuned below the resonance frequency. 
This light diffuses out, trapped by the absorption and emission process. 
Such events also shift the light's frequency to the blue on average. Eventually  the light is emitted from the dilute exterior regions of the atomic cloud, 
Fig. \ref{cooling_scheme}.  If such a scenario can be maintained it is obvious that on average the energy consumed to generate the 
blue shift in the radiation frequency will be extracted from the kinetic energy of the atoms leading to cooling.

Energy flow from the particle medium to the electromagnetic field will occur only if a single photon will go through multiple excitation/dexcitation cycles. 
This imposes a restriction on the absorption cross section and the particle density.
The particles should be stratified in the trap such that the high density is in the centre. 
Simultaneously, the particles undergo diffusional dynamics as a result of repetitive inter-particle collisions and interaction with light.
Overall, under these conditions, two different mediums, particle and photons, are captured in the trap interacting with each other by energy transfer and influencing their respective motion. We will refer to the photonic medium as the 'light medium' in the following.
The energy shift of the light is a stochastic many body effect incorporating 
multiple absorption emission cycles. It combines the asymmetry in absorption and emission line-shapes with the spectral dependence of the light trapping in the medium, the mechanism is explained in detail in the following.

\begin{figure}[htb!]
 \centering
 \includegraphics[scale=0.3]{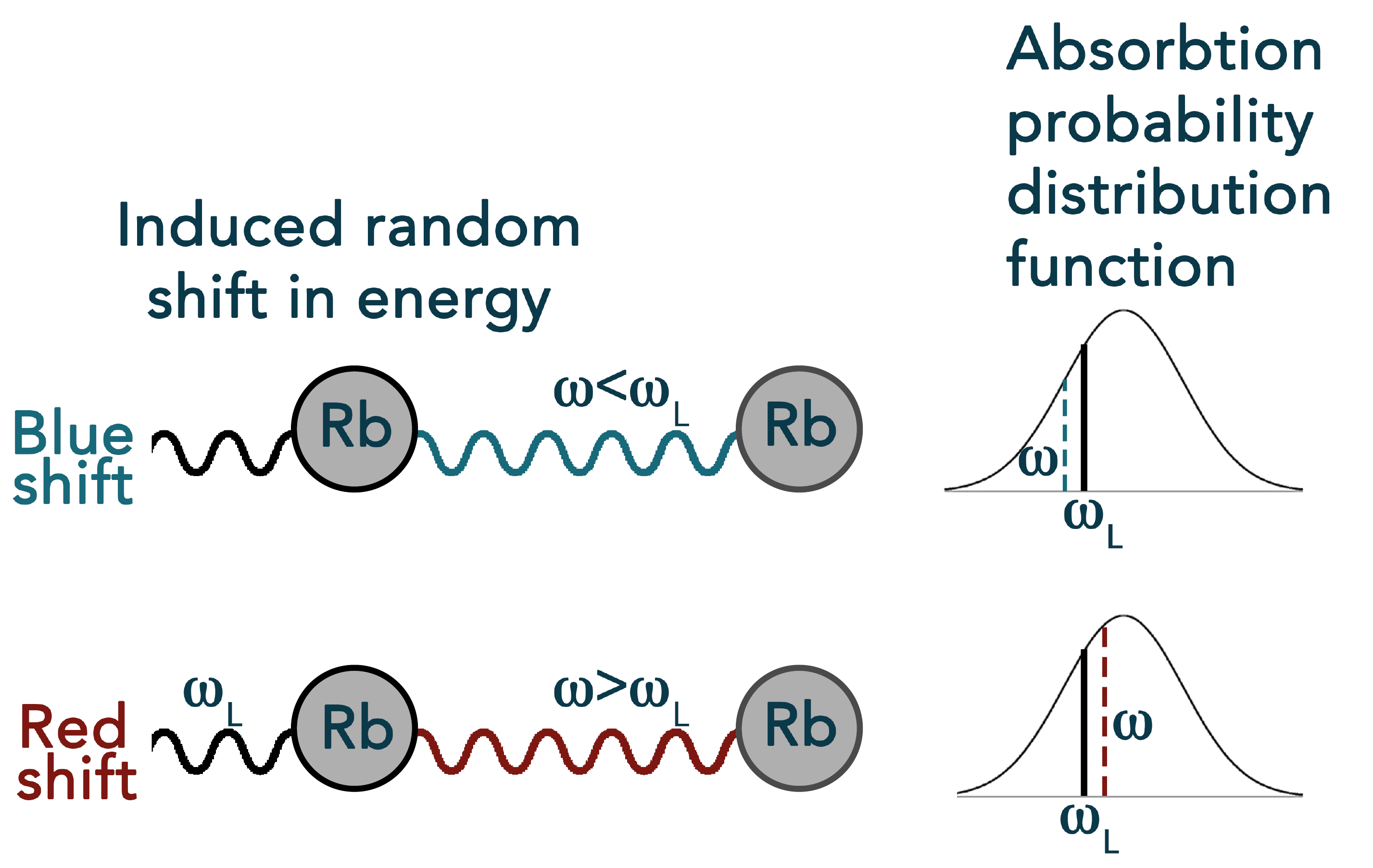} 
 \caption{  A photon of the laser frequency, $\omega_{L}$, is absorbed by a $~^{87}\mbox{Rb}$ atom. In a random process the frequency of the emitted photon, $\omega$, is shifted with respect to $\omega_{L}$. For a positive energy shift (top part) the probability to absorb increases and conversely for a negative energy shift (lower part).}
\label{figuretwo} 
\end{figure}

The photons captured in the trap undergo repetitive excitation cycles, for each cycle there is a probability of an energy shift to the photon on account of the particle's kinetic energy. Due to the asymmetry of the absorption probability function (incident light is detuned below the atomic line)  a 'blue' shifted photon will have a higher probability of being reabsorbed, Fig. \ref{figure_absorption_function}, \ref{figuretwo}. While a 'red' detuned photon probability to be reabsorbed is decreased, i.e, photons which transfer energy to the particle medium will diffuse faster through the particle medium until reaching low particle densities on the edge of the trap and escaping. This causes an effective cut off for energy transfer from the light medium to the particles. Alternatively, photons detuned to the 'blue' which reduce the particles' energy, undergo more excitation cycles, allowing further energy transfer from the particle medium to the light, Fig. \ref{figuretwo}.  The collective effect of dual dependence induces a net energy transfer between the two media and efficient cooling.

Another characteristic of the dual dependence involves pressure broadening \cite{pichler1983diffuse}: 
When the density of an atomic ensemble increases, a red shift to the atom transition frequency occurs. 
This red shift is universal and is caused by the larger polarizability of the excited state which enhances the long range attractive van der Waals force.
The particle density profile can therefore induce a spatially varying optimal absorption frequency. For a particle density which decreases radially, the resonance frequency increases accordingly. As a result, a photon starting  in the centre when shifted to the blue and propagating outward  will be reabsorbed due to new resonance conditions. 
This process will be repeated until a blue photon will escape the trap at the low density outer region.

An important issue is how to stratify the light in the opaque particle medium. A few options can be envisioned. When the trap forces are large causing
a large gradient in density a far red detuned incident light will only be absorbed at the centre. 
Once the light frequency is shifted to the blue it becomes  trapped by the particles. The only escape route is from the dilute outer boundary. 
Another option is to use electromagnetic induced transparency  EIT to inject the light to the interior.

\subsection{Energy transfer mechanism:}

\begin{figure}[htb!]
 \centering
 \includegraphics[scale=0.4]{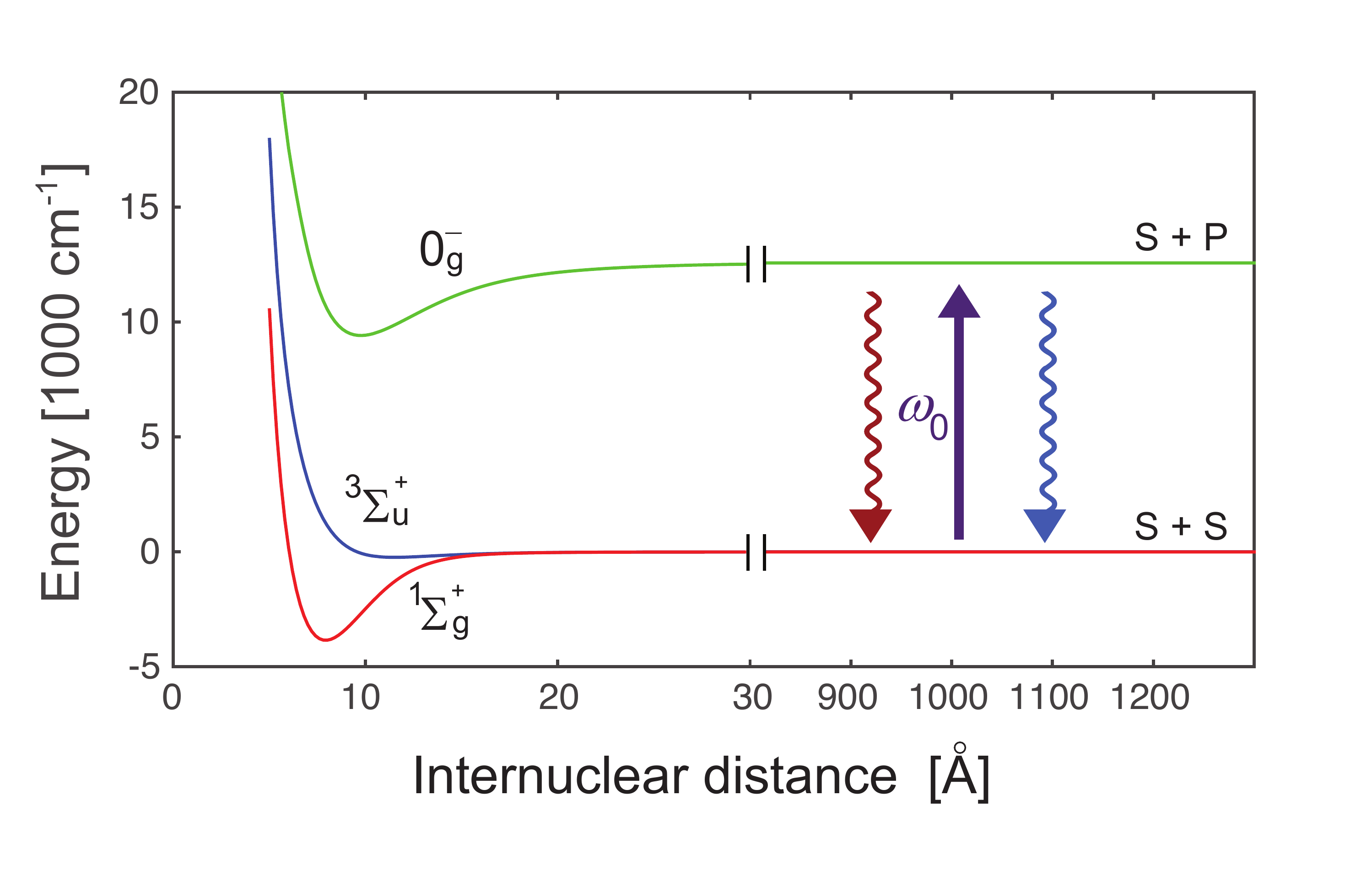} 
 \caption{\label{fig: rubidium energy levels} $~^{87}\mbox{Rb}$  energy levels, lowest singlet ground state $X^{1}\Sigma_{g}$ and  $a^{3}\Sigma_{g}$ and one of the excited states $0_{u}^{-}$.  The excitations occur in the long range part of the potential $r\approx 10^{3}${\AA}. (Energy in units of wavenumbers).}
\end{figure}

For a homogeneous atomic gas with the typical S$\rightarrow$ P transition, the excited inter-atomic van der Waals potential scales as $C_3/r^3$
compared to $C_6/r^6$ for the ground state potential, Cf. Fig. \ref{fig: rubidium energy levels}, \cite{stwalley1978pure}.
As a result, the absorption frequency varies with the relative distance. Once a photon is absorbed, the atom spends an average lifetime ($27.7\cdot10^{-8} \text{s}$)  in the excited state. At this stage, the two neighbouring atoms will undergo random relative motion, until their decay by photo-emission. This random motion, accompanied by Doppler phenomena, collisions, random electromagnetic fields and the natural linewidth of the excited state, cause an energy shift of the emitted photon.  Summarizing the phenomena: For sufficient density, the relative motion causes a change in the van der Waals potential energy at the expense of the emitted photon energy. Following the description, a random energy shift requires random relative motion between neighbouring atoms, this is indeed the case in the semiclassical limit, where the relative motion is isotropic. 

It is important to note that in the present modeling we neglected  elastic Rayleigh scattering which occurs in addition to the repeated absorption/emission events. In typical cases, such as propagation of light through biological tissue or planetary atmosphere,  \cite{chandrasekhar2013radiative, dudko2005photon}, elastic scattering constitutes the main contribution, thus influencing photon propagation. However, for photons near atomic absorption line propagating in a particle medium with a large absorption cross section, the absorption phenomena is the major contribution to the light propagation, arising from a large difference in the typical lifetime of the two processes.

In the exposition the inter-atomic interactions are modeled by 
the four electronic energy states of Rb$_2$ molecules, the two ground states; the singlet and triplet, $X\,^{1}\Sigma_{g}^{+}$  and $a\,^{3}\Sigma_{u}^{+}$  correspondingly, and $0_{g}^{-}$  and $1_{g}^{-}$  excited states \cite{steck2001rubidium}. The two ground states differ at close range distances, but for large distances ($r >100 $ {\AA}), 
the ground states' singlet and triplet coalesce, scaling as van der Waals interactions $\propto -1/r^6$. 
The excited state's long range potential scale as $\propto-1/{r^{3}}$ due to a degeneracy  of the  P state \cite{stwalley1978pure}.

\section{Modelling methods for the combined particle and light media}
\label{sec:Modeling}

\subsection{Probabilistic analysis over phase space}
The Stochastic Cooling theory was modelled by a probabilistic simulation where the physics is embedded in terms of the dynamics of continuous probability distribution functions (PDF) over phase space. This is the suitable description for diffusional behaviour and dominant collective effects. 
Both the light and particles are confined in the trap, and are described by the position and momentum  in the trap. It is important to note that for the particles the momentum is proportional to the velocity  while for the photons it is linearly dependent on the energy.

A full stochastic model involves a 12 dimensional probability function, including all the particle and photon degrees of freedom (DOF). Such a system is computationally very demanding. However, if an isotropic environment is assumed, all the axes are degenerate, and only 4 DOF (position and momentum DOF for particles and light) are required. A 4D model is still computationally challenging with respect to the desired accuracy. A solution to this problem is achieved by comparing typical time scales characterizing both media. The particle diffusion and thermalization rate is much faster than the resonant photon diffusion rate. Such a separation of time scales effectively  decouples the two ensembles in a short time regime. This assumption allows us to break down the general model to two separate phase space distributions, viz. to particle and light media. This separation follows the mean field approximation.

\subsubsection{The Fokker Planck Eq. particle ensemble}

\paragraph{Initialization:}
In the initial stage, the particles, described by the  PDF, $P$, are confined in a trap with an initial temperature $T_{init}$. $P$ is propagated in time by the Smoluchowski equation \cite{einstein1905molekularkinetischen, Sutherland1905diffusion,  von1906kinetischen, kolmogoroff1931analytischen} until steady state is reached.

\begin{eqnarray}
\label{eq:fk1}
\pd{P\brac{{x_{par},p_{par}}}}t=-\pd{}x_{par}\left({\f{p_{par}}mP} \right)+\pd{}p_{par}\left({\brac{V_{h.o}'\brac{x_{par}}+\mu_{par}p_{par}}P} \right)\\
\nonumber
+D_{par}\left({\mu_{par},T_{init}}\right)\ppd P{p_{par}}
\end{eqnarray}
where: $x_{par}$  and $p_{par}$  are the particle position and velocity, correspondingly; $m$  is the particle mass; $\mu_{par}$  is the drag constant, calculated from the experimental relaxation time \cite{bloch2001sympathetic}. $D_{par}\brac{\mu_{par},T_{init}}$ is the particle momentum diffusion function, dependent on the drag constant and temperature.

The first term on the R.H.S. describes the coupling between the velocity and location of the atoms. 
The collisions between the atoms transfer momentum between the two particles, creating an overall diffusion in momentum which is described by the last term. 
Balancing the diffusion is the  trap's potential, associated to the term 
$\frac{\partial}{\partial p_{par}}\left({V_{h.o}'\brac{x_{par}}P}\right)$, 
a mixed term coupling the confining force and the momentum and $\frac{\partial}{\partial p_{par}}\left({\mu_{par}v_{par}P}\right)$ 
is the drag term originating from particle collisions. 
An additional term can be added in extremely low temperatures, where the particle de Broglie wavelength is in the order of the mean inter-atomic 
distance and the particles' scattering length should be considered. Large scattering length should lead to
an additional spatial diffusion term. For the studied temperature regime this effect is negligible. 
For particles in an harmonic trap potential the steady state distribution has been shown to be a Gaussian with a variance dependent on the ratio between the diffusion constant and the drag force, $D_{par}=\mu_{par}k_{B}T$, where $k_{B}$  is the Boltzmann constant \cite{einstein1905molekularkinetischen, smoluchowski1952theory}

\paragraph{Coupling of the light to the particle medium:}
The interacting particle light ensemble is modeled by the  diffusion function,
$D_{par\, coupled}$. The change in the diffusion variable arises from an average energy flow from the particles' ensemble to the radiation field and momentum alterations by photon absorption/emission processes. The variable $D_{par\, coupled}$  is described in detail in Section \ref{subsec: Derivation}.

The radiation characteristics are described in detail in table \ref{Appendix: Pulse}. 

\subsubsection{The Fokker Planck Eq. for light}
\label{RTAassumptions}

To describe the light medium we construct a second 2D probability distribution function over phase space. The function is propagated in time with a F-P Eq. derived from the 'Radiative Transfer Equation' (RTE) \cite{chandrasekhar2013radiative}, similar to the Photon Diffusion Eq. \cite{ furutsu1994diffusion, durduran1997does}, further details are given in the Appendix. These equations usually describe light propagation in a scattering medium. While, in our study, we treat excitation cycles as scattering events characterized by long interaction times, resulting from the atomic decay time.

The dynamics of the light phase space are described by the following equation;

\begin{eqnarray}
\label{eq:fk2}
\pd{\phi\brac{x,p_{l},t}}t=\pd{}x\left({D_{x_{l}}\brac{\rho_{par}\brac x,E_{photon}\brac{p_{l}}}\pd{}x\phi\brac{x_{l},p_{l},t}} \right)
\\ 
\nonumber
+\pd{}{p_{l}}\left({D_{p_{l}}\brac{\rho_{par}\brac x,E_{photon},T_{par}}\pd{}{p_{l}}\phi\brac{x,p_{l},t}}\right)
\end{eqnarray}
$x$  and $p_{l}$  are the position and momentum of the photon ensemble, $E_{photon}$  is the photon energy, $T_{par}$ and $\rho_{par}$ are the instantaneous particle temperature and density, respectively.

The equation has two diffusion terms, in space and momentum, describing the diffusion in the particle medium and energy transfer. This is a similar equation to the general Photon Diffusion equation \cite{ furutsu1994diffusion, durduran1997does} but lacks any source or sink term, due to the fact that for atoms there are no clear non-radiative processes. Loss mechanisms, such as photo-association are also negligible for this case.

\paragraph{Diffusion functions explanation:}
$D_{x_{l}}\brac{\rho_{par}\brac x,E_{photon}}$  is the light position diffusion function; the photons' propagation in the particle medium is described by a diffusional movement, caused by repeated absorption/emission cycles and the isotropic nature of spontaneous emission. The particle density sets the mean distance between consecutive absorptions, the photons' energy compared to the transition line determines the probability of absorption, both affecting the diffusion rate directly.

$D_{p_{l}}\brac{\rho_{par}\brac x,E_{photon},T_{par}}$ is the light momentum diffusion function: The diffusion rate is determined by atom-atom interactions influenced by the atomic density and velocity at temperature $T_{par}$. 

The physics and interaction between both media is embedded in the properties of these variables as a function of the different parameters. A complete  analysis follows. 

\subsection{Derivation of the diffusion variables} 
\label{subsec: Derivation}
The light-matter momentum diffusion function, $D_{par\, coupled}$;

\begin{equation}
D_{par\, coupled}=m\cdot\mu\brac{k_{B}T_{inter}+E_{recoil}\cdot R }
\end{equation}

$D_{par\, coupled}$  is determined by accounting for all the different effects influencing the particles' energy or momentum, considering energy and momentum conservation. 
Each term represents a different mechanism for energy transfer. The $\mu\cdot E_{recoil}\cdot R$  term arises from the condition of a pressure balance between radiation pressure and the particles' momentum, when the recoil temperature is achieved \cite{walker1990collective}. The light medium exerts a constant radiation pressure on the particles by continuous absorption. 
The effect is insignificant when the magnetic force of the trap is bigger than the radiation pressure force, 
but should be considered when the particles are cooled to a temperature where both forces are on the same scale. $E_{recoil}$  is the recoil energy and $R$  is a constant dependent on the ratio of photons to particles or the intensity of the light. 

The energy conservation between the ensembles is described by the drag constant times the typical kinetic energy of a single particle, $\mu k_{b}T_{int}$.
The total energy of both media is kept constant by adjusting the temperature variable, $T_{inter}$. The local energy transfer between the particles to the radiation field is calculated from the net energy change due to the interactions, as well as the total energy change arising from the photon flow in and out of the trap. 

A further term $\brac{-D_{int}}$  can be added to the diffusion coefficient. The added term relates the instantaneous momentum transfer between the two media. However, the additional term is negligible in the long range due to momentum transfer accounted for in the energy transfer term, $m\cdot\mu k_{B}T_{inter}$.
\par
The present description does not account for quantum effects. At lower temperatures the theory should be modified by adjusting the modeling parameters, taking into account the particle wave characteristics. This can be done by adding a position diffusional term describing the weak localization due to collision \cite{diosi1993calderia}.

\subsubsection{ Spatial diffusion amplitude, $D_{x_{l}}\brac{\rho_{par}\brac x,E_{photon}\brac{p_{l}}}$} 

The RTE derivation for photon propagation in a highly scattering medium predicts the value of : $D_{x_{l}}=\f v{3\mu_{s}'}$  where $\mu_{s}'=\mu_{s}\brac{1-g}$  , and $\mu_{s}^{-1}$  is the mean distance between consecutive scattering events in the original derivation. In the case above, where scattering events are neglected, $\mu_{s}^{-1}$  is the mean distance between consecutive absorption events, $g$  is the scattering anisotropy constant $\mean{cos\brac{\theta}}$, which vanishes for isotropic scattering \cite{graaff2000diffusion}. Following the assumptions mentioned, \ref{RTAassumptions}, we derive a similar expression for $D_{x_{l}}$. 

\begin{equation}
D_{x_{l}}=\f{l^{2}}{3\delta t}
\label{Light spatial diffusion variable}
\end{equation}

For the range of densities common for a MOT, the emission decay time, $\delta t$, is the relevant time scale between adjacent excitations, and $l$  is the mean distance between consecutive absorptions. For near resonance light, $\delta t$ is the lifetime of the excited state, independent of the detuning \cite{guerin2017light,lagendijk1996resonant,van1999multiple} 

We assume a homogeneous medium for a small element in space. In such a medium the probability distribution for a photon 
to cover a distance $y$  without being absorbed by a particle is: 

\begin{equation}
P\brac y=\sigma_{abs}ne^{-\rho_{par}\brac x\sigma_{abs}y}
\label{eq.3}
\end{equation}
$\sigma_{abs}\brac{\nu}$  is the absorption cross section, $\nu$  is the photon frequency and $x$  the position in the MOT.
The mean free path is given by $l=\mean y=\int_{0}^{\infty}yP\brac y dy=\f 1{\rho_{par}\sigma_{abs}}$  \cite{valeau2006use}. 
From Eq. \ref{Light spatial diffusion variable} and \ref{eq.3} we obtain:

\begin{equation}
D_{x_{l}}=\left[{3\brac{\rho_{par}\brac x\sigma_{abs}\brac{\nu}}^{2}\delta t}\right]^{-1} 
\end{equation}

\subsubsection{Momentum diffusion function, $D_{p_{l}}\brac{\rho_{par}\brac x,E_{photon},T_{par}}$}
\label{subsec: Momentum diffusion variable}

The diffusion coefficient of light can be decomposed into a product of two contributions: 
1. The probability function of a photon being absorbed by a pair of interacting atoms, $G_{abs}\brac{\rho_{par},\nu}$.
2. The diffusion function describing the diffusion rate in momentum ($\propto E$), caused by the random energy shift of the absorbed photon, 
$\mathcal{D}\brac{\rho_{par},T_{par}}$.

Assuming the absorption and energy transfer mechanisms are independent and the diffusion function can be written as;
\begin{equation}
D_{p_{l}}\left({\rho_{par}\brac x,E_{photon},T_{par}}\right)=G_{abs}\brac{\rho_{par},\nu}\cdot\mathcal{D}\brac{\rho_{par},T} 
\label{eq: Light momentum diffusion variable}
\end{equation}

\paragraph{Absorption probability function:}

\begin{figure}[htb!]
 \centering
 \includegraphics[scale=0.3]{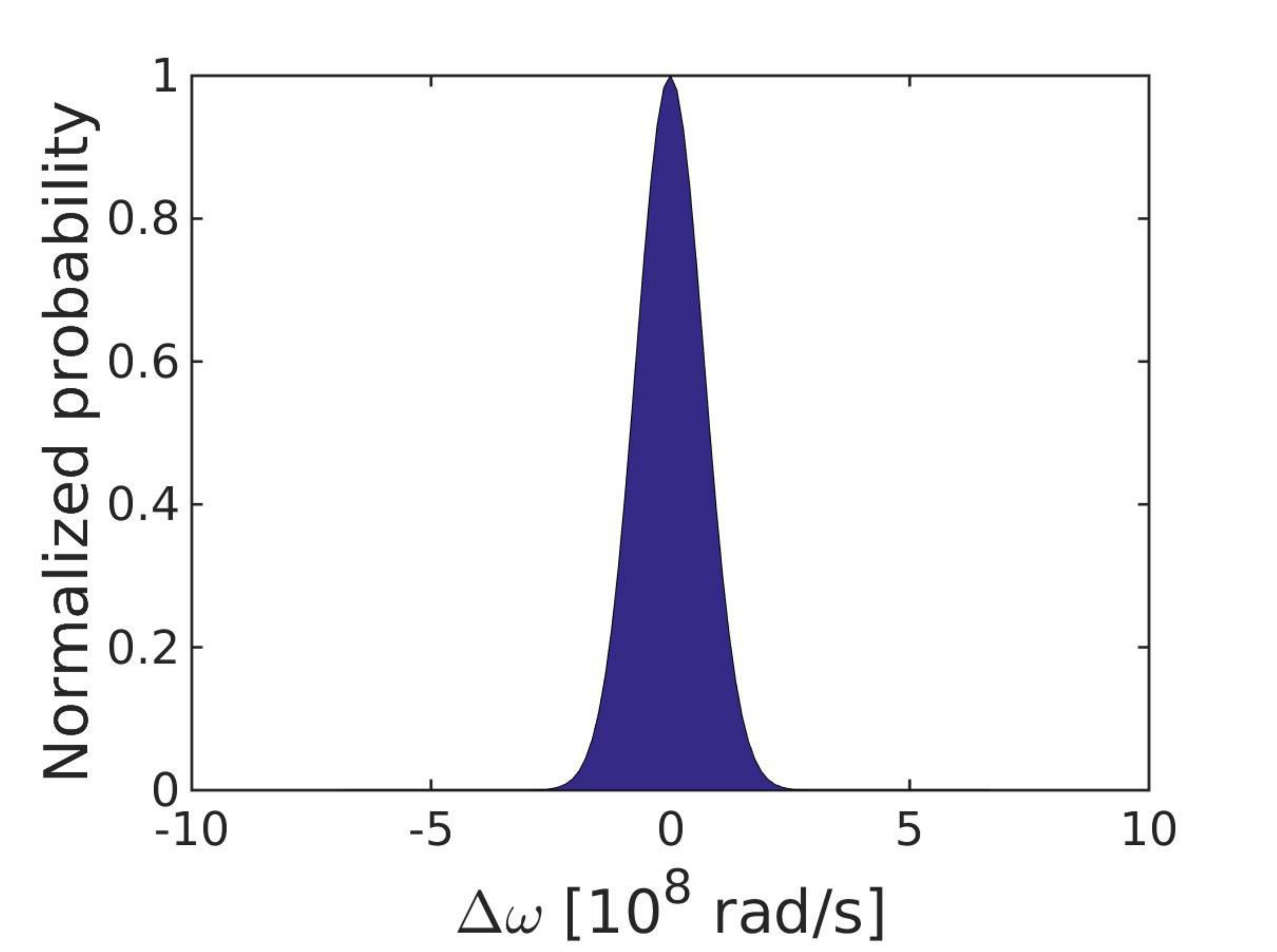} 
 \caption{The normalized absorption probability of $~^{87}\text{Rb}_2 $ as a function of the shift from the atomic transition line, $G_{abs}\brac{\rho_{par},\nu}$. 
 The details of the calculation can be found in Appendix \ref{Appendix: Absorption probabilty function}.
 The function is used as an input for the momentum diffusion function, $D_{p_{l}}$, using the experimental cross-section value to rescale the probability function.}
 \label{figure_absorption_function} 
 \end{figure}

The details of the derivation of the absorption probability function, $G_{abs}\brac{\rho_{par},\nu}$, is shown in the appendix. Here, we present an overall description and the results, Fig. \ref{figure_absorption_function}.

The absorption probability is calculated employing  a  quantum description of the absorption and emission process. 
For a low particle density, Cf Table \ref{table: Model parameters},   the analysis can be restricted to a two particle interaction.
The calculation is then reduced to a three-body interaction, two neutral 
$~^{87}\mbox{Rb}$  atoms and a light field characterizing a single photon. 
The absorption cross section is solved assuming a weak field by the time dependent perturbation theory. 
The free propagation is obtained by solving the time dependent  (TD) Schr\"odinger equation. 
The wave propagation is calculated using the Chebychev polynomial expansion method with a Fourier grid, 
together with a Gaussian Random Phase approach \cite{tal1984accurate, koch2009two}. The details of the propagation are described in Appendix \ref{table:numerics}.

\paragraph{Energy transfer between the atom and radiation field:}

There are a number of processes which cause a photonic energy shift: The natural line  broadening due to spontaneous emission 
\cite{tannoudji1973mecanique,weisskopf1930calculation}, and Doppler phenomena \cite{siegman1986lasers}. However, in high density the most dominant phenomena is pressure broadening and pressure shift \cite{rotondaro1997collisional, peach1981theory}, 
which arise from atom-atom interactions. The process is stochastic and can be viewed as a 1D random walk on an energy axis. In such a case the momentum diffusion amplitude is the squared mean momentum related to the energy shift per unit time. 
The function can be calculated as, 
$\mathcal{D}\brac{\Delta p}\propto\f{var\brac{F\brac{\Delta E}}}{\delta t}$, where $var\brac{F\brac{\Delta E}}$  is the variance of a specific probability function, $F\brac{\Delta E}$. $F\brac{\Delta E}$  describes the probability for a certain energy shift, between the emitted and absorbed photons. We show here the main point of the derivation of $F\brac{\Delta E}$. 

We begin with change in energy due in a typical excitation:
\begin{equation}
\Delta E=-\f{C_{3}}{r_{f}^{3}}+\f{C_{6}}{r_{f}^{6}}+\f{C_{3}}{r_{i}^{3}}-\f{C_{6}}{r_{i}^{6}}=C_{3}\bigg ({\f 1{r_{i}^{3}}  -\f 1{r_{f}^{3}}}\bigg )+C_{6}\bigg ( \f 1{r_{f}^{6}}-\f 1{r_{i}^{6}} \bigg ) 
\end{equation}

where $r_{i}$  and $r_{f} $ are the relative distance between a pair of functions at absorption and excitation times respectively, and $C_{3}$, $C_{6}$  are the van der Waals potential's constants.

Transforming to the centre of mass and relative velocity coordinates, the velocity distribution is a Maxwell Boltzmann distribution of particles with a reduced mass $\mu=m/2$  and kinetic energy of $E_{k}=\f{p_{r}^{2}}{2\mu}$: 
\begin{equation}
f\brac v=\sqrt{\f{\mu}{2\pi k_{B}T}}e^{-\f{\mu v^{2}}{2k_{B}T}} 
\end{equation}

The initial relative density determines the average distance, $r_{i}=\brac{\rho\brac {x}}^{-1/3}$,($x$ is the position in the trap), and the final distance is written in terms of the decay time and the initial distance, $r_{f}=r_{i}+v\cdot\delta t$, where $r_{i}\gg v\cdot\delta t$  in the relevant density and temperature range.

By expanding up to the first term in $v\cdot\delta t$, we obtain a relation between the energy transferred and the particle's relative velocity. From this relation the energy transfer distribution function is obtained by a random variable transformation for the Maxwell Boltzmann distribution.

\begin{equation}
\Delta E=3\brac{\rho\brac x}^{4/3}\brac{C_{3}-2\rho\brac xC_{6}}\cdot v\cdot\delta t=C\brac{x} v
\label{Taylor expansion of the photon energy shift}
\end{equation}

\begin{equation}
C\brac{x}=3\brac{\rho\brac x}^{4/3}\delta t\brac{C_{3}-2\rho\brac xC_{6}} \nonumber
\end{equation}

The distribution function of the change in energy for a single excitation;

\begin{equation}
F\brac{\Delta E}=N_{norm}e^{-\f{\mu\brac{\Delta E}^{2}}{2C^{2}k_{B}T}}  \nonumber
\label{Distribution of the change in energy for a single excitation}
\end{equation}

Making an ansatz in equation \ref{eq: Light momentum diffusion variable}, the radiation field momentum diffusion amplitude can be written as;  

\begin{eqnarray}
D_{p_{l}}\brac{\rho_{par}\brac {x},E_{photon},T_{par}}=G_{abs}\brac{\rho_{par},\nu}
\\
\nonumber
\cdot\f{\sbrac{3\brac{\rho\brac x}^{4/3}\brac{C_{3}-2\rho\brac xC_{6}}}^{2}\delta t\cdot k_{B}T_{par}}{\mu_{par}\cdot c^{2}} 
\end{eqnarray}
where $c$  is the speed of light.

The photon diffusion function is highly dependent on the density of the particles and the spatial distribution of photons in the trap. The linear temperature dependence demonstrates the fact that when the particles cool it becomes harder to extract entropy.

\begin{figure}[htb!]
 \centering
 \includegraphics[scale=0.3]{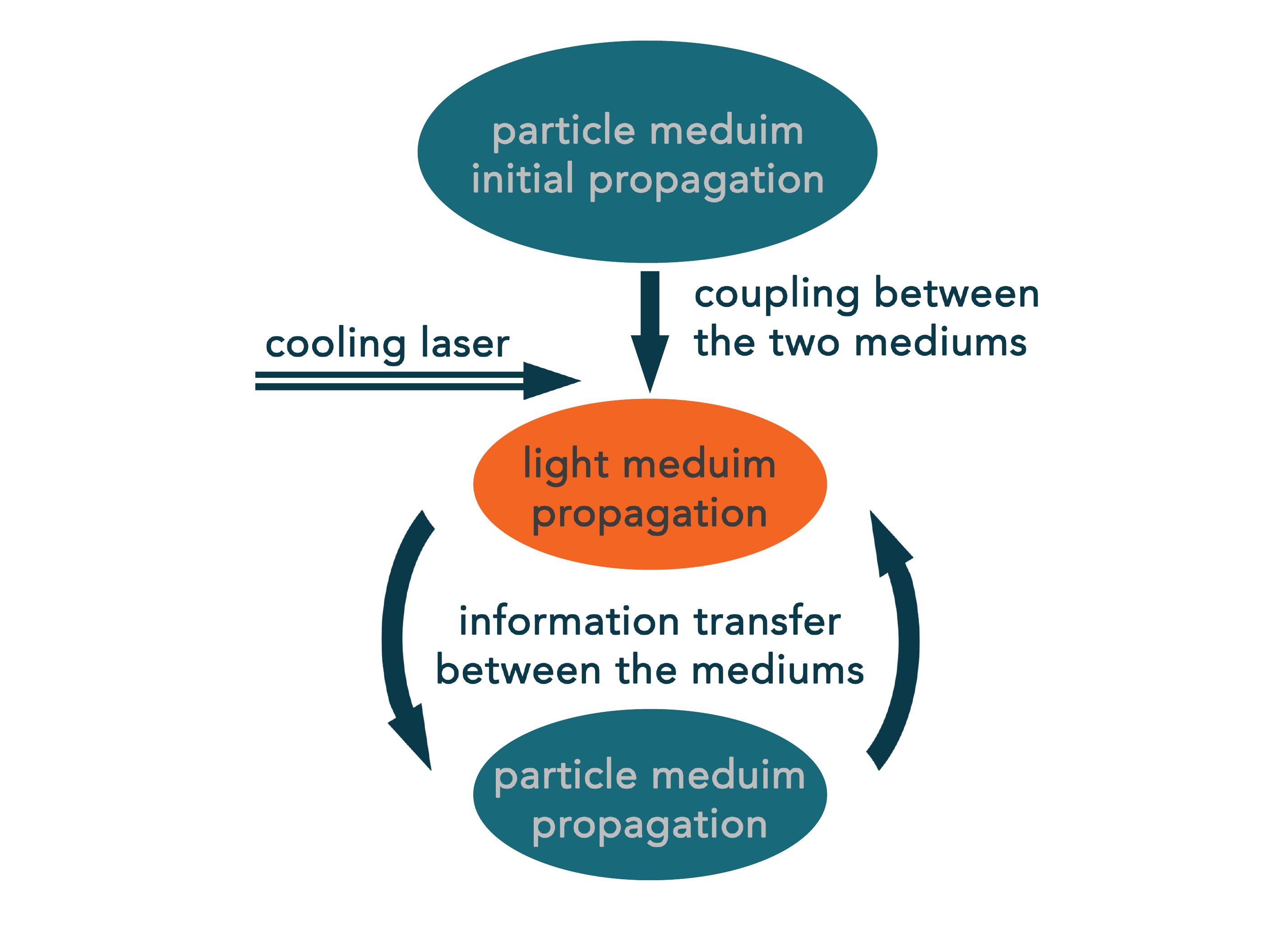} 
 \caption{A a schematic flow chart of the modelling method, as described in section \ref{Final_modelling_summary}}
 \label{flow_chart} 
 \end{figure}

\subsection{Final modelling summary}
\label{Final_modelling_summary}

The Fokker-Plank equation is propagated by a Chebychev polynomial expansion method for the evolution operator $U\brac{t+\Delta t}=e^{-\hat{G}\Delta t}\rho\brac t$, where $\rho\brac t$  is the modelled distribution function at time $t$  and the propagator, $\hat{G}=\pd{\rho\brac t}t$, is the corresponding Fokker Planck operator \cite{kosloff1994propagation,  johnson2011notes}. A Fourier method is used to calculate the derivative terms in $\hat{G}$ operation. This scheme is highly accurate and efficient. The two-phase space distribution of light and particle ensembles are propagated simultaneously for small time laps, transferring information about energy, momentum and density between the models, Fig. \ref{flow_chart}. Absorbing boundary conditions are applied to the light density function to account for the photons escaping the trap. In addition, new photons are added with a frequency distribution corresponding to the laser source. 
Such a scenario models a constant laser incident intensity.

\section{Results A: Probabilistic analysis of the Stochastic Cooling}
\label{sec:Results A}

\begin{figure}[htb!]
\center{
 \includegraphics[scale=0.5]{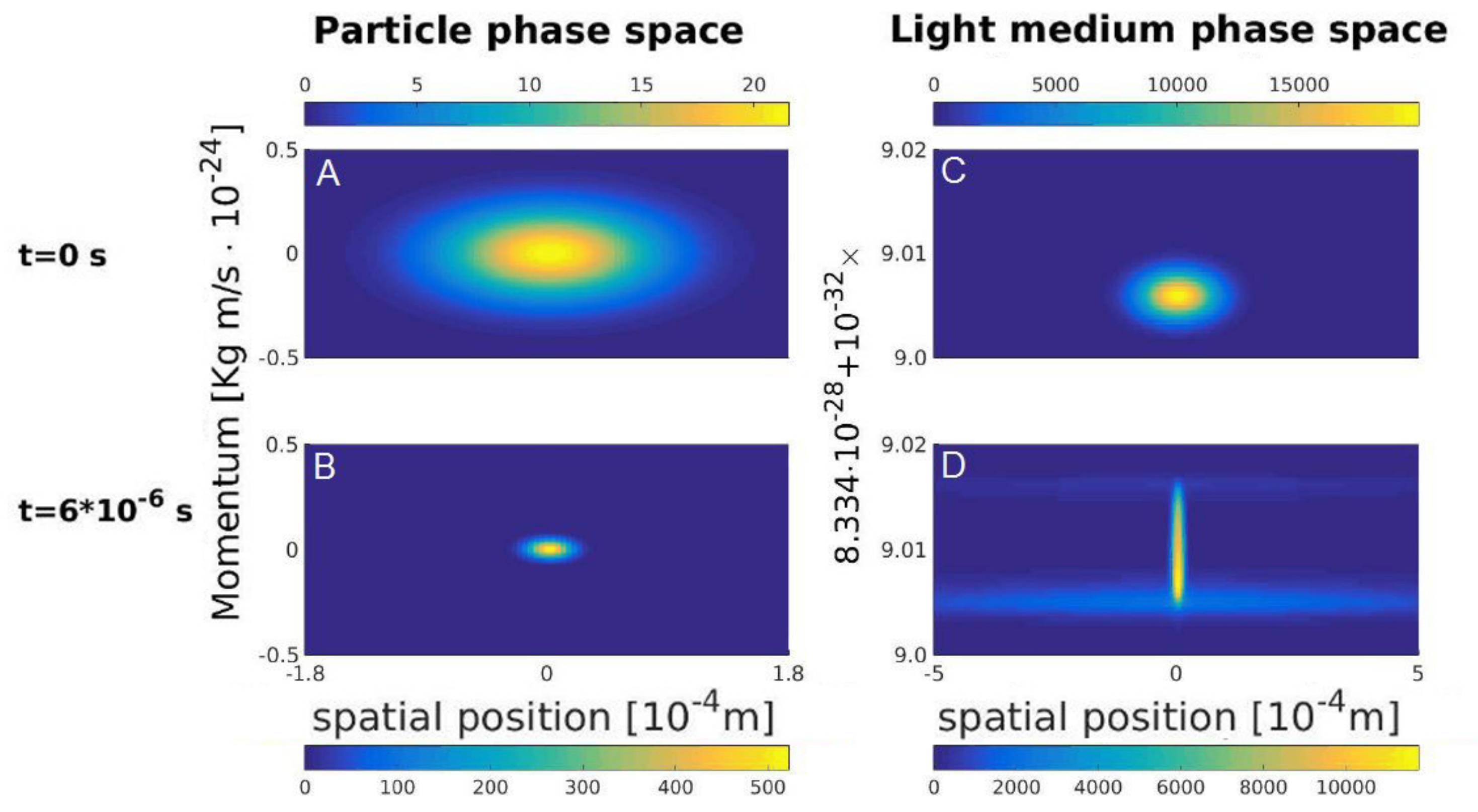}}
 \caption{ The particle phase space on the left (A,B) before coupling to the radiation field (A) and after at time $t\approx 6*10^{-6} \text{s}$ at $T=10^{-4} \text{K}$ (B). 
 The right hand side represents the light phase space before the coupling (C) and at time $t$ (D). 
 The vertical axis of all the figures describes the momentum, and the relevant scaling of the units is presented on the left of the axis.}
\label{fig:1} 
\end{figure}

Following the evolution of the initial particle Gaussian distribution after a transient time its phase space distribution is compressed. This is a signature of cooling. Figure \ref{fig:1}, (A,B plots) shows an increase of phase space density after $6\mu \text{s}$. On the other hand, the light medium experiences a fast  broadening of the momentum distribution (time scale of $0.1 \mu \text{s}$). For low momentum, the distribution reaches a threshold due to a rapid decrease of the absorption probability and a fast spatial diffusion of the low (red detuned) momentum photons escaping the trap. 
While high momentum photons are confined for longer periods of time in the particle medium.  This occurs until photons reach off-resonant frequencies, leading to a fast diffusion for extremely high frequency blue detuned photons. The phenomena can be seen in figure \ref{fig:1} D  as off-resonance photons (large gaps between the resonant momentum, $8.334901e^{-28} \frac{\text{Kgm}}{\text{s}}$; bright horizontal strips in the figure) diffuse rapidly. At this stage of the process the cooling continues and approaches a constant rate, (see Fig. \ref{fig:2}), resulting from continuous replacement of the diffused photons by laser light absorbed by the particle medium.

The cooling rate is linear for the initial coupling with light but eventually saturates because 
the energy transfer depends on the particle velocity,  $D_{p_{l}}\propto T_{par}$. 
The cooling rate slows down for low temperatures until reaching the quantum regime 
where additional processes should be incorporated in the model. 
An example of such an effect is a further contribution to the spatial diffusion resulting from localization 
of the particle wave packet due to collision with a neighbouring particle.

\subsection{Comparison of different trap potentials}
\label{sec:subsection1}

\begin{figure}[htb!]
 \center{
 \includegraphics[scale=0.3]{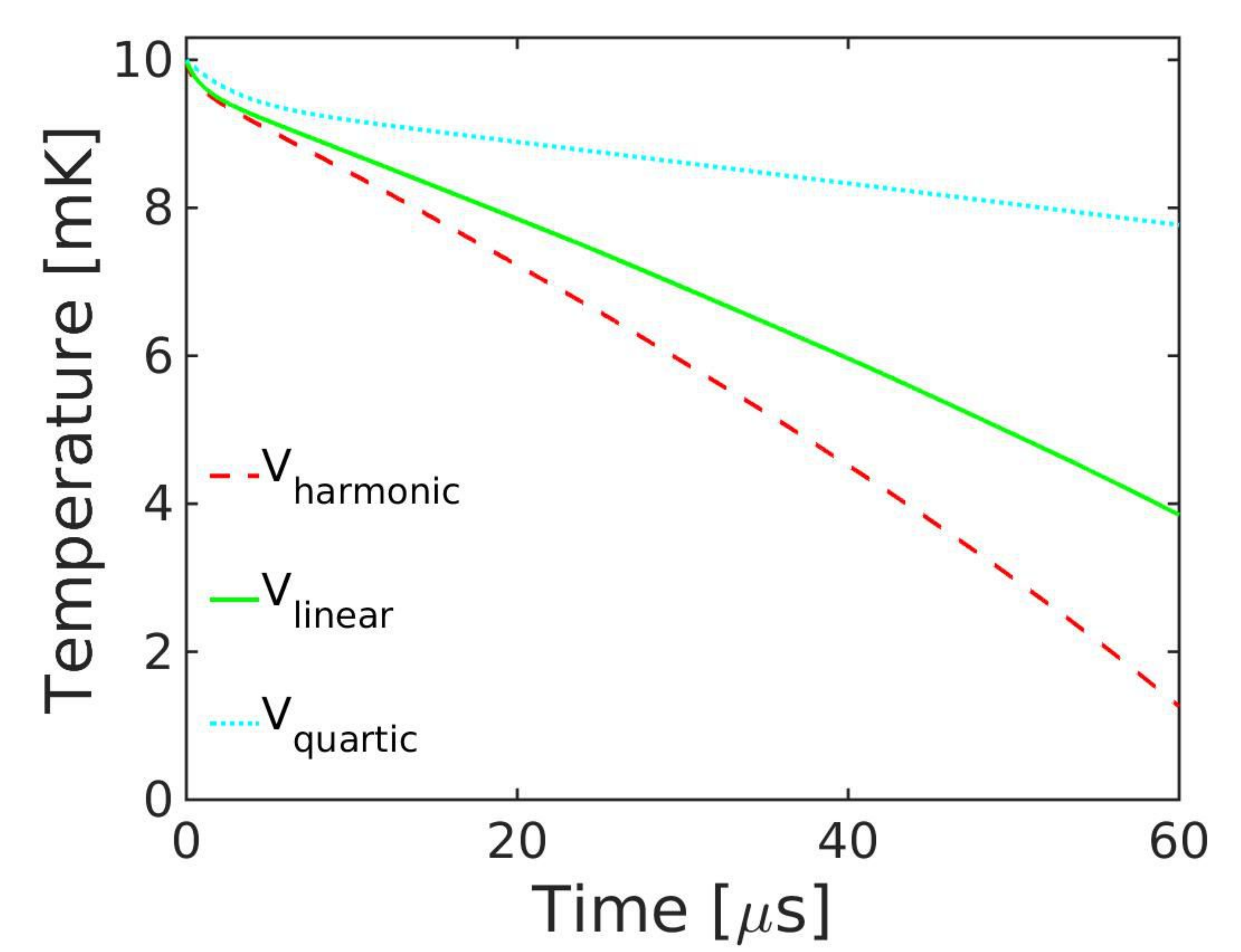}}
 \caption{Particle temperature as a function of time, for different potentials. For a density of $\rho=10^{14} \text{cm}^{-3}$; The trap potential: $V_{harmonic} = \frac 12 kx^{2}$ ; $V_{linear}=k|x|$; $V_{quadratic}=\frac 12 kx^{4}$.}
 \label{fig:temp_vs_time}
 \label{fig:2} 
 \end{figure}

The trap's potential shape determines the particle density, which in turn affects the probability of a photon to be absorbed by the particle medium. 
To test the cooling sensitivity a set of models were studied with different potentials:  harmonic, linear, and quartic potentials. 
Comparison between different trap shapes was made while keeping the potential energy at the positions $x=\f{L}{4}$, $x=\f{3L}{4}$, almost identical.
In addition, an equal  amount of particles was used for both simulations. The results are presented in Figure \ref{fig:temp_vs_time}. 
The most efficient cooling rate is predicted by the harmonic trap,$1.45\cdot10^{2} \text{K/s}$,  
which is almost by $50\%$ larger than the particle cooling in a linear trap, $1.02\cdot10^{2} \text{K/s}$. 
The quartic $\brac{\propto x^{4}}$ potential shows a cooling rate of  $37.1 \text{K/s}$.
Due to the red shift of absorption with density the cooling is optimal when there is a significant gradient in particle density such as in the harmonic trap.

\subsection{Comparison of different densities}
\label{sec:subsection2}

\begin{figure}[htb!]
 \centering
 \includegraphics[scale=0.3]{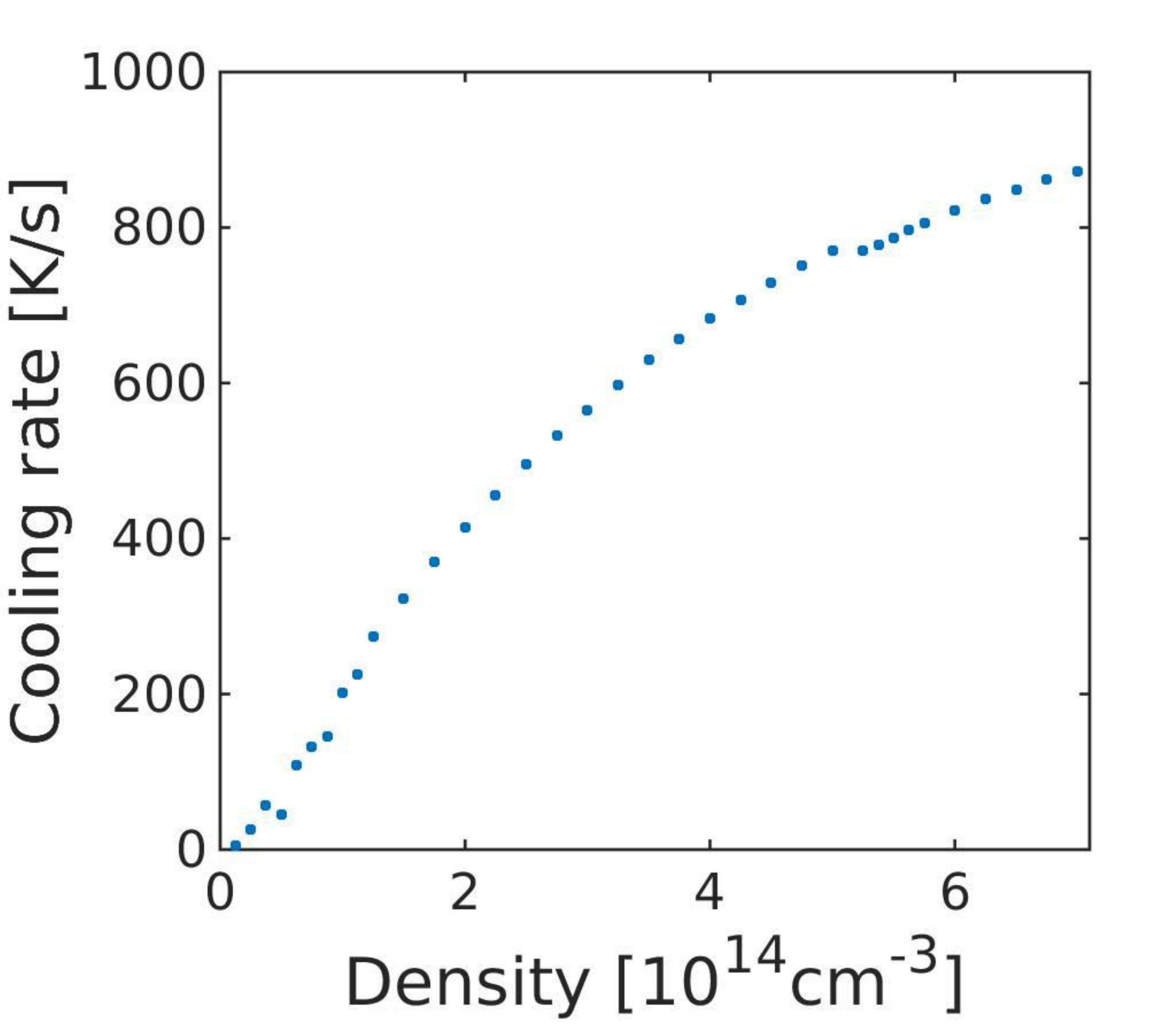}
 \caption{ 
 The cooling rate in absolute value as a function of the initial average particle density.}
 
 %%%The file forming the plot is cooling_rate_vs_density in absorbing_boandery_condition folder%
  \label{fig:cooling_rate_vs_density}
\end{figure}

A direct connection between the average density in the trap and the cooling rate was found.
At low density a linear increase in the cooling rate is observed Cf. Fig. \ref{fig:cooling_rate_vs_density}. 
At higher densities the  the cooling rate reduces.
This is in a density range which is considerably lower in comparison with the quantum regime. 
In such a regime, the basic many body cooling phenomena should still be valid but complimented by quantum corrections to the model.      

\break
An estimation of the maximum cooling rate is obtained by considering a sphere filled with a uniform gas of  atoms with density $\rho$, and a typical spontaneous emission lifetime of $\tau$. The bound for the cooling rate can be calculated by noticing that only the atoms  at the outer boundary of the sphere effectively emit energy. 
The number of atoms per a unit area occupying the outer shell is $N_{surf}\sim\rho^{2/3}$. On average the excited atoms out of $N_{surf}$ will emit a blue shifted photon giving an energy difference of $\hbar\Delta\omega$, with a rate of $\dot{Q}=\frac{\hbar\Delta\omega}{\tau}$. Furthermore, if $1\%$ of the atoms on the surface are excited at a certain instant, the upper bound to the cooling rate  can be estimated as:
\begin{equation}
R\sim0.01\cdot\dot{Q}=0.01\frac{\hbar\Delta\omega}{k_{B}\tau}
\label{r}
\end{equation}
Inserting in equation \ref{r} the data for $~^{87}\mbox{Rb}$: $\rho=10^{13}\text{cm}^-3$, $\Delta \omega \le 10^7 \text{rad/s}$, $\tau_d = 27.7 \text{ns}$ gives $R\sim 10^6 \text{K/s}$. Where the  $\Delta \omega$
was estimated from Figure \ref{fig:cooling_rate_vs_density} considering the stochastic nature of the process.
Note that this is an upper bound, not taking into account the stochastic nature of the cooling and possible heating sources. Comparing to the current modeling, it predicts a smaller cooling rate of the order of 
$R\sim 10^3 \text{K/s}$.

The geometrical arguments can explain the dependence of the cooling rate on the density. At low density the whole volume emits therefore a linear scaling is expected as seen in Fig. \ref{fig:cooling_rate_vs_density}.
For high density asymptotic cooling rate should scale as of $\rho^{2/3}$. For the data of  the asymptotic cooling rate (Fig.  \ref{fig:cooling_rate_vs_density}), we obtain the scaling of $R\sim \rho^{0.67}$, in accordance with the geometrical analysis.

\section{Enhanced Stochastic Cooling - An extension of the Stochastic Cooling method}
\label{sec:Enhanced}

The Stochastic Cooling of $~^{87}\mbox{Rb}$ atoms, described in Section \ref{sec:Theory}, utilizes the energy gap dependence on the inter-atomic distance. To generalize this mechanism we propose a method applicable to different types of constituents. The main idea incorporates additional control of the energy gap between the ground and first excited
potential energy surface. 

The Stochastic Cooling method, (Sec. \ref{sec:Theory} and \ref{sec:Modeling}) requires a spatial dependence on the energy gap between the ground and excited states. 
$~^{87}\mbox{Rb}$  is a unique case, the ground and excited energy states scale differently with the relative distance between the atoms, Cf. Fig. \ref{fig: rubidium energy levels}, resulting in an energy gap with a sufficient spatial gradient. In the general case, both energy states scale similarly, and the spatial gradient may not be sufficient to achieve efficient cooling.

An enhancement of the spatial gradient between the energy states can be induced by employing a second CW field with a frequency , $\omega_{1}$, in resonance with the transition line between the excited state, $E_{e}$, to a higher excited state, denoted by $E_{f}$. It is crucial that the frequency $\omega_{1}$  should be different from the atomic transition, not affecting the excitation process from the ground state.

\paragraph{Derivation:}
The atomic energy gap between subsequent levels roughly scales as $\propto\left({\f 1{n^{2}}-\f 1{\brac{n+1}^{2}}}\right)$, demonstrating that the first energy gap is much bigger than the other gaps. The big difference between the energy gaps allows an explicit treatment of a two-level-system coupled to an oscillating classical field \cite{cohen1977quantum}. The solution is given in terms of the Rabi frequency , $\Omega$, linearly dependent on the vector electric field amplitude of the laser, $\vec{E_{\mathcal{L}}}$. 

We will focus on two energy levels of the excited state, $\ket{e_{1}}$  and the level of a higher excited state, $\ket{e_{2}}$. The classical radiation field induces an energy shift to the bare Hamiltonian levels, and the new shifted states are given by

\begin{equation}
E_{\pm}=\pm\f{\hbar\sqrt{|\Omega|^{2}+\Delta^{2}}}2 
\end{equation}

where $\Delta=\omega_{L}-\brac{\omega_{f}-\omega_{h}}$, can be neglected for a resonant radiation, $\omega_{1}\equiv\omega_{f}-\omega_{e}\gg\Delta$. 

\begin{equation}
E_{\pm}=\pm\f{\hbar\Omega}2 
\end{equation}

For a classical radiation field, of frequency $\omega_{1}$, with a spatial dependence, the intensity varies in the trap. For example, a high intensity light focused at the centre of the trap will have a gradient toward lower intensities at the edge of the trap. The intensity gradient results in a spatial dependent Rabi frequency, $\Omega\brac x$, where $x$  is the trap's radial coordinate. Concentrating on a pair of atoms in the trap, the Rabi frequency can be written as a function of the inter-atomic distance, $r$. This leads to an excited state $E_{e}$  which varies spatially as well, while the ground state stays unperturbed by the classical EM field. This phenomena induces a spatial energy dependent gap between the ground and excited states, with a gradient depending on the EM intensity.

Once an energy gap is controlled by a tuning laser of frequency $\omega_{1}$, a second laser of frequency $\omega_{0}=\omega_{e}-\omega_{g}$  is applied to the particle medium. As in the Stochastic Cooling method, the laser of a frequency $\omega_{0}$ induces excitations between the ground and excited states. The relative random 
motion of the atoms in the excited state will induce an average energy transfer and cooling. This process is controlled by the electric field amplitude $\vec E_{\mathcal{L}}$  with a frequency $\omega_{1}$. 

\begin{figure}[htb!]
\centering
\includegraphics[scale=0.45]{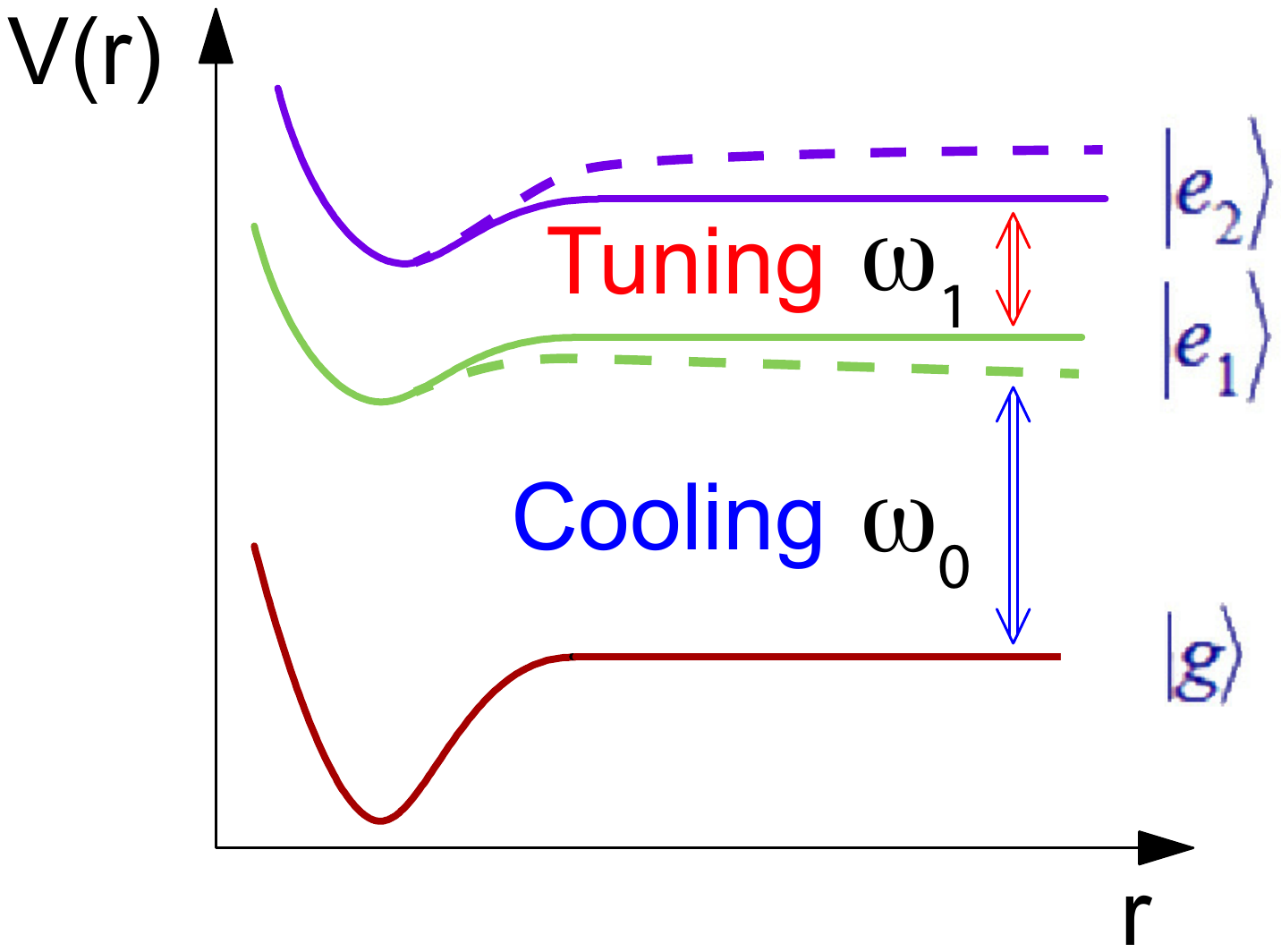} 
\caption{\label{figurefive} The excited state, $\ket{e_{1}}$  , and the next energy state, $\ket{e_{2}}$, are coupled by a tuning laser  of frequency $\omega_{1}$ generating a Stark shift
which is intensity dependent. 
The Stark shift is enhanced at resonance conditions occurring at a specific inter-atomic distances.  As a result,  the potential $\ket{e_{1}}$ is modified and with it the resonance conditions of the cooling laser $\omega_{0}=\frac{E_{e}-E_{g}}{\hbar}$.  By varying the intensity of the tuning laser in the trap we obtain a gradient in the absorption probability of the cooling laser.
 }
\end{figure}

The energy level configuration and the addition of a second tuning laser is similar to the scenario utilized for Electromagnetic Induced Transparency, (EIT). In EIT a  combined  AC-Stark splitting and quantum interference results in a transparency at a frequency of the probe (cooling) laser. Similar applications have been achieved for Rb and Pb. Such phenomena, under favourable circumstances, will allow an easy penetration of the photons to a partly opaque particle medium \cite{xiao1995measurement,field1991observation}.   

The control of the energy gap gradient enables the Enhanced Stochastic Cooling to operate at lower densities relative to the densities required for Stochastic Cooling of Rubidium. For Stochastic Cooling of $^{87}$Rb, the high densities are required because of the small dependence of the energy gap on the inter-atomic distance. Once the gradient is engineered with an external field, the energy gap can be shifted to longer inter-atomic distances. For this method, the required density is bounded only by densities  where the photons are characterized by diffusional motion.

We present in table \ref{fig:constituents} the sufficient densities for applying Enhanced Stochastic Cooling of different Alkali atoms appear .

\begin{tabular}{ |p{4cm}||p{4cm}| }
 
\hline
\label{fig:constituents}
 Constituent     & Required density $\sbrac{\text{cm}^{-3}}$  \\
 \hline
 Rubidium   &  $1.47\cdot10^{6}$      \\
 Sodium  & $2.85\cdot10^{6}$ \\ 
 Caesium &  $1.04\cdot10^{6}$   \\

 \hline
\end{tabular}

\subsection{Optimization of the tuning radiation field} 
 
In the following section we discuss how optimization of tuning laser intensity and frequency affects the cooling.
 
\subsubsection{Intensity variations of the tuning radiation field} 

The question arises what is the optimal field profile? The laser frequency is determined by the gap $\omega_{1}=\omega_{f}-\omega_{e}$, but different intensity profiles can be realized. Modern-day optics allow creating many intensity profiles, utilizing optical holographic lenses, and state-of-the-art optical devices. An optimization including all possible scenarios can be complicated. However, we notice that by a similar derivation as in section \ref{subsec: Momentum diffusion variable} the diffusion constant $Dp_{l}$ is proportionate to the square of the tuning laser's electric field gradient. Proportionality suggests that a largely varying field in the trap region will result in increased cooling to the particle media, Cf. Fig. \ref{fig:temp_vs_time_dipole_force}.

\subsubsection{Frequency variation of the classical radiation field} 

The frequency of the second laser source can be tuned to induce cooling for a specific atomic density. In the last section we consider a classical radiation field, applied to the trap, with resonance frequency $\omega_{1}$  matching the asymptotic transition $\ket{e_{1}}\ra \ket{e_{2}}$. However, the energy gap in resonance to the transition changes along the inter-atomic distance, $r$. Modern experimental techniques allow accurate control of the laser frequency and spatial intensity.
This enables  control of the exact region of the inter-atomic distances which are coupled to the cooling field. A laser with a frequency of $\omega_{1}\brac{r_{i}}$  will couple between the ground and excited states in a region near $r_{i}$, inducing a gradient in the energy gap between the states. The gradient will induce cooling, originating from a pair of atoms with a certain inter-atomic distance. Alternately, when averaging over the inter-atomic distances, the laser frequency, $\omega_{1}\brac{r_{i}}$, will match an atomic medium of a density $\rho\brac{r_{i},x}$, (where $x$  is the radial component of the trap). 

\section{Results B: Enhanced Stochastic Cooling}
\label{sec:Results B}

Enhanced Stochastic Cooling was modelled on $~^{87}\mbox{Rb}$ atoms, in a similar method as described in section \ref{sec:Modeling}. 
A second tuning laser in conjunction with the cooling laser is employed with a wavelength of $1475.6 $ nm  ($~^2P_{1/2} \rightarrow ~^2D_{3/2}$).
Once the light and particle media were coupled, the two phase spaces were propagated in time, while synchronizing the parameters after each time step.
Energy transfer is assumed to originate solely from coupling of the external field. 
The diffusion functions are recalculated based on the derivation presented in Section \ref{sec:Enhanced}. 

\begin{figure}[htb!]
 \centering
 \includegraphics[scale=0.3]{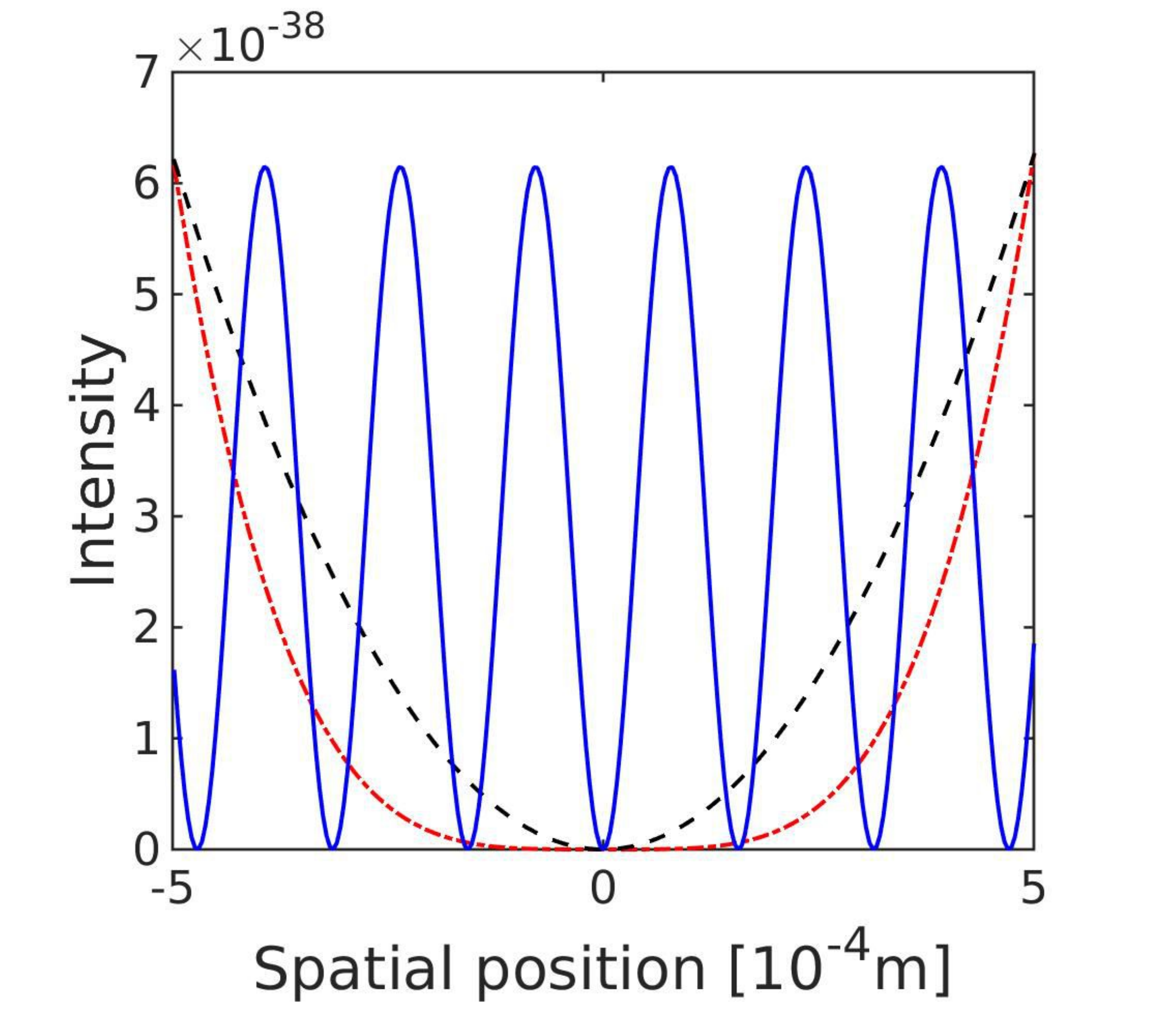}
 \caption{Profiles of tuning light intensity: Black dashed $E_{1}=2\cdot10^{14} x^2/2$, Red dash-dot $E_{2}=5\cdot10^{12} x^4 $, Blue solid $E_{3}=4.96\cdot10^{13}sin(200 x)/200$ (MKS). The field profiles where adjusted so that the maximum intensity in the trap will be equal.}
 \label{fig:Intensity_plot}
\end{figure}

\begin{figure}[htb!]
 \centering
 \includegraphics[scale=0.3]{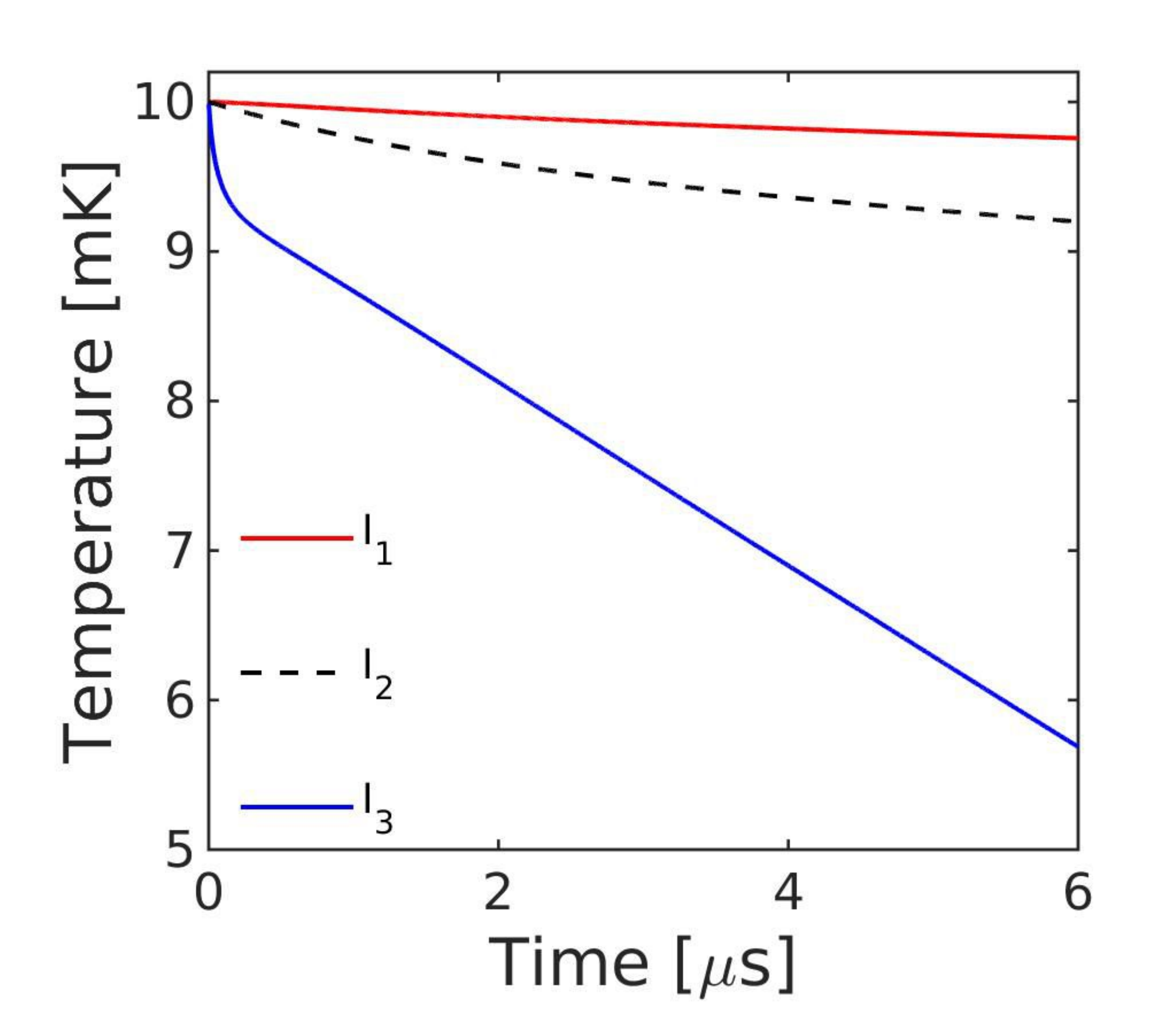}
 \caption{ Particle temperature as a function of time for different external field profiles corresponding to figure \ref{fig:Intensity_plot}.}
 \label{fig:temp_vs_time_dipole_force}
\end{figure}

A number of different intensity profiles were studied, $E_{1}\propto x^{2}$,$E_{2}\propto x^{4}$, and also a highly oscillating profile $E_{2}$. Cf Fig.
\ref{fig:Intensity_plot}, \ref{fig:temp_vs_time_dipole_force}. 
The sinusoidal profile shows the fastest cooling rate, $6.85\cdot10^{2} \text{K/s}$ as a result of a large gradient. The other profiles  $E_{2}$ and $E_{1}$ have an
inferior cooling rate of, $1.33\cdot10^{2} \text{K/s}$ and $40 \text{K/s}$, correspondingly.  

\break

\section{Discussion}

Cooling of neutral atoms via collective many-body interactions 
is an efficient universal cooling scheme,
applicable  as a complementary method to prior cooling methods, \cite{kozyryev2016radiation,dieckmann1998two},  or  independently. 
 
At sufficient particle density and large absorption cross sections photons are trapped in the particle media. In such density regimes, the photon propagation in the trap is characterized by diffusion. A single photon exhibits a large number of excitation cycles, 
allowing energy and entropy transfer between particles and photons. The mechanism proposed depends on the collective behaviour of the particles and light media, 
giving rise to coordinated dynamics.  The cooling rate depends on particle density, on the density gradient and
the asymmetry in the spectrum between absorption and emission. 
On average, an absorbed red photon will be emitted as a blue photon.
Density of similar values used in the demonstration have been achieved experimentally, \cite{camara2014scaling,ferlaino12,chomaz2017observation}, 
$\left({10^{13}- 10^{15} \text{cm}^{-3}}\right)$.
A simple experimental setup for stochastic laser cooling requires the ability to change the trap potential and with it the density. This will also allow to check
the validity of the theory.

The main thermodynamic principle unifying all  cooling methods is an increase of the total entropy of the joint particle and light ensembles. 
The energy transfer from the particle medium to the light medium decreases the entropy of the particle ensemble. This comes at the
expense of the light entropy where the constant radiation loss from the trap is the entropy generating mechanism.

Our modeling for Rubidium demonstrates that cooling can be achieved by the 'Stochastic Cooling method'. A single 'cooling laser', coupling the ground and lowest excited state ($\ket{g} = X\,^{1}\Sigma_{g}^{+}$,   $ a\,^{3}\Sigma_{u}^{+}$ ; $\ket{e} =  0_{g}^{-}$, $1_{g}^{-}$) is sufficient. Modeling  predicts an efficient cooling for a density range of $\brac{1-7} \cdot 10^{14} \text{cm}^{-3}$, cooling rates ranging between $100-800 \text{K/s}$. 
Furthermore, the model predicts that the asymptotic cooling rate for  a fixed trap volume  will scale with the density as $\rho^{2/3}$ and be linear at low density. 
This is in accordance with a simple geometrical model.

This scheme can be extended to cool other constituents by adding a second 'tuning laser' coupling the first state and a higher excited state. The generalized cooling method, 'Enhanced Stochastic Cooling' , allows universal extension of cooling for different types of neutral atoms as well as molecules. The experimental requirement is an additional CW laser.
The rate of cooling can be controlled by determining the values of the gradient of the intensity of the tuning laser, $|\pd{E_{tune}}{x}|$, \cite{merli2009photoassociation}. 
The generalized scheme predicts efficient cooling rates that can be maximized by choosing an intensity profile with a large spatial gradient in the trap, 
Cf. Fig. \ref{fig:temp_vs_time_dipole_force}. 

The phenomena, enabling cooling of Rubidium (Stochastic Cooling), is related to pressure line shift and pressure broadening, which arises at sufficiently high densities \cite{vdovic2006absorption}. The pressure shift and inter-atomic interactions allow energy transfer between particle translational degrees of freedom to internal degrees of freedom and to the photonic medium. The magnitude of the pressure shift influences the cooling rate directly. Similar effects, arising for increased densities, 
can be seen in other condensed matter phenomena, such as charge transfer to solvent 
and modifications to absorption/emission spectra for liquid phase relative to a gas phase spectrum. 
For cooling, higher particle densities increase the pressure shift and, in turn, the gradient of the energy gap between the ground and excited states. 
As a result of the stochastic nature of the process, a larger gradient leads to faster cooling.

The mechanisms described, responsible for energy transfer from the particle to the light medium, is valid for the semi-classical regime. 
At low temperatures, the present theory should be modified  by a quantum theory. 
The crossover temperature is when the de Broglie thermal wavelength is in the range of the mean particle distance ($T~>~\frac{h^{2} \rho ^{2/3}}{2\pi m k_{B}}$).
For $~^{87}\text{Rb}$ this temperature is $\sim 10^{-6} \text{K}$ for  a density of $\rho = 10^{14} \text{cm}^{-3}$.
At low temperatures the asymmetry between red and blue shift
emission is larger. The reason is that the ground state density is peaked at the attractive region
of the van der Waals potential and the emission is biased toward the outer turning point of the vibration of the excited potential which is larger.
Near the BEC limit, additional corrections can be made based on particle wave characteristics. These are out of the scope of this paper. 
The Enhanced Stochastic Cooling method  is applicable to low densities and lower temperature regimes. 
The details are presented in Table \ref{fig:constituents}.

Manipulation of cold molecules and cooling molecules to extremely low temperatures  has been one of the main focal points of the Atomic Molecular Optical research field, \cite{stuhl2012evaporative,bochinski2003phase,narevicius2009single,
zeppenfeld2012sisyphus,bethlem2000electrostatic,bethlem2003production,jochim2003bose}. The method proposed is applicable to molecular cooling experiments. For sufficient densities, light can be trapped for long time periods in the molecular medium, and an energy transfer is predicted.

In contrast to the general simplicity of  laser cooling atoms, the higher number of degrees of freedom  in molecules induces a complex internal energy structure. These features complicate the cooling process due to additional relaxation channels which leads to induced heating. 
For efficient cooling the molecules need to have large diagonal Franck-Condon factors. 
This will allow repeated electronic excitations while minimizing the excitation of vibrational states. 
Low inelastic rates are favourable  which result in heating and a fast molecular loss rate. 
In addition, the energy transitions should match the available laser cooling  frequencies.  
A number of different constituents qualify for efficient Stochastic cooling or Enhanced Stochastic cooling, ${\rm OH}$,${\rm CaF}$ and ${\rm YO}$. 
 
${\rm CaF}$ has been cooled to velocities of $\approx 10 \pm 4 \text{m/s}$, which is below the capture velocity of a molecular MOT \cite{hemmerling2016laser}, and has suitable electronic transitions. The cooling laser can be applied, detuned slightly below the $X^{2}\Sigma_{1/2}^{+}$, $v=0$, $N=-1$,   $A^{2}\Pi_{1/2}$, $v=0,~J=+3/2$ transition of $606\text{nm}$,  while the coupling laser couples between $A^{2}\Pi_{1/2}$, $v=0~,J=+3/2$ and $C^{2}\Pi _{1/2}$, $v=0,~J=-1/2$ states of $729.5 \text{nm}$. 
Additional lasers and a magnetic field may be required in order to bring back dark magnetic sub-states to the optical cycle and reduce population loss to excited vibrational states. Such a scheme is envisaged to induce Enhanced Stochastic cooling. 
A spatial gradient in the energy gap between the ground state and excited state allows efficient cooling. 
The gradient is created by the polarizability difference between the two states.
This means that a dense ensemble of ${\rm CaF}$ trapped in a MOT could be cooled further, towards sub-millikelvin temperatures.  

Other molecular candidates are ${\rm OH}$ radical and ${\rm YO}$ which have been confined in a trap \cite{sawyer2007magnetoelectrostatic,stuhl2012evaporative,hummon20132d} and studied in context with optical cooling \cite{yeo2015rotational,collopy2015prospects,bochinski2003phase}. 
They have a suitable internal energy structure with convenient optical transitions in the visible light. 
Both show similarities to $\rm CaF$, allowing application of Stochastic Cooling methods. 
Similarly to the case of $\rm CaF$, additional pumping lasers and a magnetic field could be needed to ensure a closed cooling cycle.
These additional lasers prevent population trapping in dark states by re-pumping back to the cooling cycle.

Almost all molecules are more polarizable in the excited state. As a result, a gradient in the energy gap will arise and the Stochastic Cooling method is therefore applicable since the Stochastic Cooling method is based on van der Waals forces. Molecules with higher polarizability could be cooled more efficiently.  The molecular density should be in the range where two-body
elastic collisions are dominant over three-body inelastic ones.
To conclude, 'Enhanced Stochastic Cooling' can greatly increase the class of atoms and molecules that can be cooled to sub-Kelvin temperatures.

\section{Acknowledgements} 
\label{sec:Acknowledgements}
We thank Amikam Levi and
Shimshon Kallush for fruitful discussions and
Maya Dann and Moshe Armon for their help. This work was partially supported by the ISF - Israeli Science Foundation

\section{Appendix}
\label{sec: Appendix}

\subsection{Rubidium data table} 
D1$\,\mbox{transition }\brac{5^{2}S_{1/2}\ra5^{2}P_{1/2}}$ \cite{steck2001rubidium}

\label{table: Rb data table}
\begin{tabular}{ |p{5cm}||p{4cm}| }
\hline
\label{fig:constituents}
 Wavelength (Vacuum)     & 794.979 nm  \\
 \hline
 Lifetime   &  27.7 ns      \\
 \hline
 Recoil Energy &  22.8236 kHz    \\
  \hline
 Effective Far-Detuned Saturation Intensity & $4.484 \text{mW/cm}^{2}$  \\
  \hline
 Effective Far-Detuned Resonant Cross section  & $1.082\cdot10^{-9} \text{cm}^{2}$  \\

 \hline
\end{tabular}

\label{table: Model parameters}
\subsection{Model parameters}

\label{table: Rb data table}
 \begin{tabular}{ |p{5cm}||p{5cm}| }
\hline
\label{fig:constituents}
  Density Range   & $10^{13}-10^{14}$ $\text{cm}^{-3}$  \\
 \hline
 Trap Length   &  1 mm      \\
 \hline
  Cooling Laser Frequency &  $3.7711\cdot10^{14} \text{s}^-1 $    \\
  \hline
 Particle Photon Ratio in The Trap & 1  \\
  \hline
  Initial Temperature & $0.01 \text{K}$  \\
  \hline
 \end{tabular}
 
 \subsection{Pulse Parameters, Cf Sec. \ref{subsec: Momentum diffusion variable}}

 \begin{tabular}{ |p{4cm}||p{4cm}| }
\hline
\label{Appendix: Pulse}
 Variance   &  $10^{-16} \text{s}^{2}$      \\
 \hline
 Amplitude &  Normalized so the cross section will fit the experimental value    \\
 \hline
\end{tabular}

\subsection{Numerical methods} 
\label{table:numerics}

The dynamical equations to be solved for the particles and  light Eq. \ref{eq:fk1} and \ref{eq:fk2} have the structure:
\begin{equation}
\label{eq:difer}
\frac{\partial}{\partial t} F(x,p,t) = {\bf  O} F(x,p,t)
\end{equation}
where $F(x,p,t)$ is the probability function in phase space and ${\bf O}$ is a differential operator defined in Eq.  \ref{eq:fk1} and \ref{eq:fk2} .
We can write a formal solution for a short time step $\Delta t$:
\begin{equation}
\label{eq:prop}
F(x,p,t+\Delta t) \approx e^{{\bf O} \Delta t} F(x,p,t)
\end{equation}
The exponent in Eq. \ref{eq:prop} is expanded by a Chebychev polynomial of order $N$ \cite{kosloff1994propagation}:
\begin{equation}
e^{{\bf O} \Delta t} \approx \sum_{k=0}^{N} C_k (\Delta t) T_k ( {\bf O} )
\end{equation}
where $C_k$ are expansion coefficients (Bessel functions) and $T_k(x)$ is the Chebychev polynomial of order $k$.

The following table presents the details of the Fourier-Chebychev numerical scheme for solving coupled Fokker-Planck equations.\\
\par
\begin{tabular}{|l||l|}
\hline
\label{fig:constituents}
Grid size & $10^{-3}$ m\\
\hline
 Number of spatial grid points  in position & 250  \\
 Number of spatial grid points  in momentum & 500  \\
 \hline
 Grid spacing   &  $\Delta r = 4 \cdot 10^{-6}$ m    \\
 Grid spacing   &  $\Delta p = 4.8491\cdot 10^{-27}$ kg~m/s    \\
 \hline
 Order of Chebychev Polynomial (light)& 45  \\
 \hline
 Grid size & $2\cdot 10^{-3}$ m\\
 \hline
 Grid points of light medium position& 500   \\
 Grid points of light medium momentum& 100  \\
  \hline
 Grid spacing  (light) &  $\Delta r = 4 \cdot 10^{-6}$ m    \\
 Grid spacing  (light)  &  $\Delta p = 4.8395 \cdot 10^{-36}$ kg~m/s    \\
 \hline
 Typical time step (light)& $10^{-9}$ s  \\
  \hline
  Number of time steps & 6000\\
 \hline
 Order of Chebychev Polynomial (light)& 271  \\
 \hline
\end{tabular}

\label{table: numerical parameters}

\subsection{Estimate the number of excitations for a single photon}

The photon propagation through the atomic medium can be modeled as a 3D random walk, resulting from repeated absorption/emission cycles. 
The square of the distance that a photon reaches after N steps, or variance is;
\begin{equation}
var_{3d}=N\varepsilon^{2}    
\end{equation}

where $N$ is the number of absorption/emission cycles and $\varepsilon$ is the length of each step between consecutive absorption events. Assuming a spherical trap with a uniform density; 

\begin{equation}
\varepsilon=P_{abs}^{-1}\cdot\rho^{-\f 13}    
\end{equation}
 
where $P_{abs}=\rho^{\f 23}\sigma$ is the absorption probability. (For $\rho^{\f 23}\sigma\leq1$ the equality holds.)

To escape the trap, the photon has to reach a distance of $R$ (trap radius) from the center of the trap.
\begin{equation}
N=\f{R^{2}}{\left({P_{abs}^{-1}\cdot\rho^{-\f 13}}\right)^{2}}=R^{2}\rho^{-2}\sigma^{-2}
\end{equation}

\begin{equation}
\rho\brac{N}=\left({NR^{-2}\sigma^{2}}\right)^{-\f 12}    
\end{equation}

\subsection{Absorption probability function}
\label{Appendix: Absorption probabilty function}

We solve the transition probability between the ground and excited state of a quantum system of two  $~^{87}\mbox{Rb}$  atoms and a light field characterizing a single photon. 

The original Hamiltonian, with no coupling to a radiation field is:

\begin{equation}
\hat{H}_{g/e}=\hat{T}+\hat{V}_{g/e}=\f{\hat{P}}{2m}+V_{g/e}\brac{\mathbf{r}} 
\end{equation}
In the presence of an electromagnetic field the two surfaces of the ground and excited states are coupled by the interaction of the field and the dipole momentum operator. 
The new Hamiltonian is written as:
\begin{equation}
\label{eq: The full Hamiltonian}
\hat{H}=\hat{H_{g}}\otimes\hat{P_{-}}+\hat{H_{e}}\otimes\hat{P_{+}}+\varepsilon\brac t\hat{\mu}\otimes\hat{S}_{+}+\varepsilon\brac t\hat{\mu}\otimes\hat{S}_{-}=\bigg[\begin{array}{cc}
\hat{H_{e}} & \varepsilon\brac t\hat{\mu}\\
\varepsilon*\brac t\hat{\mu} & \hat{H_{g}}
\end{array} 
\bigg]
\end{equation}
The electromagnetic field is given by;
\begin{equation}
\varepsilon\brac t=\bar{\varepsilon}\brac te^{-i\omega_{L}t}+\bar{\varepsilon}*\brac te^{i\omega_{L}t} 
\end{equation}
where $\omega_{L}$ is the laser carrier frequency and $\bar{\varepsilon}\brac t$ the envelope of the pulse.
After the rotating wave approximation, the Hamiltonian reduces to;
\begin{equation}
\label{RWAhamiltonian}
\hat{H_{S}}=\bigg [ \begin{array}{cc}
\hat{H_{e}}-\hbar\omega_{L}/2 & \bar{\varepsilon}\brac t\hat{\mu}\\
\bar{\varepsilon}\brac t\hat{\mu} & \hat{H_{g}}+\hbar\omega_{L}/2
\end{array}
\bigg ] 
\end{equation}
The amplitude of absorption of a photon is calculated considering a system following dynamics governed by $\hat{H}_{S}$. 
For the basis states $\brac{\{\psi_{k}\}}$  of $\hat{H}_{S}$  , the amplitude transfer from the eigenstate $\ket{\psi_{i}}$  to eigenstate $\ket{\psi_{n}}$, 
after time $t$, is given by the time dependent perturbation theory. Assuming a weak field the amplitude, to good approximation, is given by the first order term;
\begin{equation}
\label{eq: time pertubation amplitude}
b_{n}^{\brac 1}\brac t=-\f i{\hbar}\int_{0}^{t}{e^{i\omega_{ni}t}\hat{W_{ni}}\left({t'}\right)dt} 
\end{equation}
where $\hat{W}_{ni}\brac t$  is the time perturbation term; $\omega_{ni}=\f{E_{n}-E_{i}}{\hbar}$.
Defining
\begin{equation}
c_{n}\brac t=b_{n}\brac te^{-iE_{n}t/\hbar} 
\end{equation}
With the help of Eq. \ref{eq: time pertubation amplitude}

\begin{equation}
c_{n}\brac t=-\f i{\hbar}\int_{0}^{t}{d\tau e^{-iE_{n}\brac{t-\tau}/\hbar}\hat{W_{ni}}\brac{t'}e^{-iE_{i}\tau/\hbar}} 
\end{equation}
For a perturbation $\hat{W_{ni}}\brac{t'}=\bar{\varepsilon}\brac t\hat{\mu}$,  and the following identity; $e^{-iE_{n}\brac{t-\tau}/\hbar}\ket{\psi_{n}}=e^{-i\hat{H_{n}}\brac{t-\tau}/\hbar}\ket{\psi_{n}}$ 

\begin{equation}
c_{n}\brac t=-\f i{\hbar}\int_{0}^{t}{d\tau e^{-iE_{n}\brac{t-\tau}/\hbar}\bar{\varepsilon}\brac t\hat{\mu}\brac{t'}e^{-iE_{i}\tau/\hbar}} 
\end{equation}
Similarly, for the excited state, using Eq. \ref{RWAhamiltonian}.

\begin{equation}
\ket{\psi_{e}\brac t}=-\f i{\hbar}e^{-\f i{\hbar}\brac{\hat{H}_{e}-\hbar\omega_{L}/2}t}\int_{0}^{t}{d\tau e^{\f i{\hbar}\hat{H_{e}}\tau}e^{-i\hbar\omega_{L}\tau}\hat{\mu}\bar{\varepsilon}\brac{\tau}e^{-\f i{\hbar}\hat{H_{g}}\tau}\ket{\psi_{g}\brac 0}}
\end{equation}
Assuming a narrow Gaussian pulse which centred at $t=0$  far away from the source, the integral boundaries can be taken to infinity.
\begin{equation}
\bar{\varepsilon}\brac{\tau}=\f B{\sqrt{2\pi\sigma_{t}^{2}}}e^{-\f{\tau^{2}}{2\sigma_{t}^{2}}} 
\end{equation}
where $B$  is an amplitude constant.
Using the identity \cite{ashkenazi1997quantum} (Sec. \ref{sec:Modeling}) 
\begin{equation}
\sigma_{A}\brac{\omega_{L}}\propto\bra{\psi_{i}}{ \hat{A}}{\ket{\psi_{i}}}\ra\hat{A}=\hat{\mu}\infint d\tau e^{\f i{\hbar}\brac{\hat{H_{e}}-\hat{H_{g}}-\hbar\omega_{L}}\tau}\hat{\mu} 
\end{equation}
Assuming the system is in the ground state at the initial time, the propagator is given by;
\begin{equation}
\hat{A}=\hat{\mu}\infint d\tau\bar{\varepsilon}\brac{\tau}e^{\f i{\hbar}\brac{\hat{H_{e}}-\hbar\omega_{L}}\tau}e^{-\f i{\hbar}E_{g}\tau}\hat{\mu}
\end{equation}
When the transition dipole moment is constant in postion and momentum, the solution of the integral gives;
\begin{equation}
\hat{A}=\hat{\mu}^{2}Be^{-\f{\sigma_{t}^{2}\Delta^{2}}2}
\end{equation}
where $\Delta=\f 1{\hbar}\hat{H_{e}}-\omega_{g}-\omega_{L}$ 
Defining, $\alpha=\f{\sigma_{t}^{2}}2$, as well as emitting the global phase from the expression and defining $c=\brac{\omega_{g}+\omega_{L}}$ and decomposing the Hamiltonian to the kinetic and potential terms, $H_{e}=\hat{T}+\hat{V}_{e}$. The expression for the propagator is given by;

\begin{equation}
\sigma_{A}\propto {\langle \psi_{g}|}{e^{-\f{\sigma_{t}^{2}}2\brac{\hat{G}+\hat{F}+\hat{K}}}|\psi_{g}\rangle}
\end{equation}

where:

\begin{equation}
\hat{G}=\hat{T}^{2}-2\hbar c\hat{T}
\end{equation}

\begin{equation}
\hat{F}=\hat{V_{e}}^{2}-2\hbar c\hat{V_{e}}
\end{equation}

\begin{equation}
\hat{K}=\hat{T}\hat{V_{e}}+\hat{V_{e}}\hat{T}
\end{equation}

Using the Zassenhaus formula to expand the exponent, we find that the high order commutators can be neglected \cite{magnus1954exponential}.

Taking the first term in the expansion, the cross section can be summarized by the expression;

\begin{equation}
\sigma_{A}\propto {\langle \psi_{g}|}{e^{-\f{\sigma_{t}^{2}}2\hat{G}}e^{-\f{\sigma_{t}^{2}}2\hat{K}}e^{-\f{\sigma_{t}^{2}}2\hat{F}}|\psi_{g}\rangle}
\end{equation}

Where $\hat{G}$ and $\hat{F}$  are the kinetic and potential energy terms correspondingly, and $\hat{K}$  is a correlation term.

The solution is a Gaussian function with a variance of $2.89$ $\brac{\text{nHz}}^2$, Cf. Table \ref{Appendix: Pulse}, centred around $\brac{V_{e}-V_{g}}/\hbar$, for large $r$'s the contribution of the van Der Waals interactions to the probability to be absorbed is negligible. However, for a short range the interaction will shift the resonance frequency towards lower frequencies in comparison with the atomic transition line, influencing the optimized detuning from resonance, $\Delta\omega_{optimise}$, used for optimal cooling.

\paragraph{Direct cross section calculation:}
We can decompose the initial thermal state to random phase Gaussian wave functions, when assuming a contribution of the kinetic term only. The approximation is valid for the density and temperature regime in our experiment, Cf. Table \ref{table: Model parameters}, where the ground state potential has a minor effect on the wave function of the ground state.

\begin{equation}
e^{-\f{\hat{H}}{k_{B}T}}\approx e^{-\f{p^{2}}{2mk_{B}T}} 
\end{equation}

Each thermal Gaussian wave function, in the momentum representation, has a temperature dependent standard deviation, $\sigma=\sqrt{mk_{B}T}$  and an added random phase $G\brac p=e^{-\f{p^{2}}{2mk_{B}T}+ipR_{0}}$. 

In the position representation this amounts to an ensemble of Gaussians centred at different locations $\{R_{0}\}$. In the final stage of the calculation all Gaussians are summed and averaged, the random phases cancel one another constructing the asymptotic thermal state propagated in time.

The overall effect of the described calculation is equivalent to the following process: Each Gaussian, centred at a different location, is coupled to an electric field at time $\tau$, the EM field couples the ground and excited states resulting in a population transfer to the excited state. The excited state is then propagated until time $t$  to achieve a single realization. The overall excited state is then achieved by integrating on all possible transition times, $\tau$. The calculation converges to the first order time perturbation term assuming a weak pulse. 

This process is repeated for different laser frequency shifts, $\Delta\omega$, and an absorption probability distribution function dependent on the laser frequency shift is achieved.

\subsection{Relation between the cross section and the matrix element}

Deriving the proportionality $\sigma_{A}\brac{\omega_{L}}\propto{\langle \psi_{i}|}{\hat{A}|\psi_{i}\rangle}$.
The power can be written as %\textcolor{reference to hellman fyenmann}; 
\begin{equation}
P=\f{dE}{dt}=\mean{\f{dH}{dt}} 
\end{equation}
Making an ansatz of Eq. \ref{eq: The full Hamiltonian} 
\begin{equation}
P=\mean{\f{d\varepsilon\brac t}{dt}\hat{\mu}\otimes\hat{S}_{+}+\f{d\varepsilon^{*}\brac t}{dt}\hat{\mu}\otimes\hat{S}_{-}}=\mean{\f{d\varepsilon\brac t}{dt}\hat{\mu}\otimes\ket{\psi_{e}}\bra{\psi_{g}}+\f{d\varepsilon ^{*}\brac t}{dt}\hat{\mu}\otimes\ket{\psi_{g}}\bra{\psi_{e}}}
\end{equation}
Inserting the density matrix expression; $\rho=\f 12\brac{\ket{\psi_{g}}\bra{\psi_{g}}+\ket{\psi_{e}}\bra{\psi_{e}}}$ : 
The state is written as: 

\begin{eqnarray*}
\f{d\varepsilon\brac t}{dt}\bra{\psi_{e}}\hat{\mu}\ket{\psi_{g}}+\f{d\varepsilon^{*}\brac t}{dt}\bra{\psi_{g}}\hat{\mu}\otimes\ket{\psi_{e}}
 \\
=-2\text{Real}\left({\f{d\varepsilon\brac t}{dt}\bra{\psi_{e}}\hat{\mu}\ket{\psi_{g}}}\right)=-2\text{Real}\left({\f{d\varepsilon\brac t}{dt}\mean{\hat{\mu}\otimes\hat{S}_{+}}}\right) 
\end{eqnarray*}
The power at time $t$ ;

\begin{equation}
\label{eq: power 1}
P\brac t=-2\text{Real}\left({\f{d\varepsilon\brac t}{dt}\bra{\psi_{e}\brac t}\hat{\mu}\ket{\psi_{g}\brac t}}\right)\propto\bra{\psi_{g}\brac t}\hat{\mu}\infint d\tau e^{\f i{\hbar}\brac{\hat{H_{e}}-
\hat{H_{g}}-\hbar\omega_{L}}\tau}\hat{\mu}\ket{\psi_{g}\brac t} 
\end{equation}
The power is proportionate to the population change which has a linear dependency on the cross section
\begin{equation}
\label{eq: power 2}
P=\hbar\omega_{0}\f{dN_{e}}{dt}\propto\sigma\brac{\omega_{L}} 
\end{equation}
combining Eq. \ref{eq: power 1} and \ref{eq: power 2} we get the desired relation;
\begin{equation}
\sigma_{A}\brac{\omega_{L}}\propto\bra{\psi_{g}\brac t}\hat{\mu}\int_{-\infty}^{\infty} d\tau e^{\f i{\hbar}\brac{\hat{H_{e}}-\hat{H_{g}}-\hbar\omega_{L}}\tau}\hat{\mu}\ket{\psi_{g}\brac t}=\bra{\psi_{g}\brac t}\hat{A}\ket{\psi_{g}\brac t}
\end{equation}

\subsection{Energy transfer between the atom and radiation field and calculation of $\mathcal{D}\brac{\rho_{par},T_{par}}$} 
The energy change due to a typical excitation is:
\begin{equation}
\Delta E=-\f{C_{3}}{r_{f}^{3}}+\f{C_{6}}{r_{f}^{6}}+\f{C_{3}}{r_{i}^{3}}-\f{C_{6}}{r_{i}^{6}}=C_{3}\left({\f 1{r_{i}^{3}}-\f 1{r_{f}^{3}}}\right)+C_{6}\left({\f 1{r_{f}^{6}}-\f 1{r_{i}^{6}}}\right) 
\end{equation}
where $r_{i}$  is the inter-atomic distance for time $t$  when the photon is absorbed and $r_{f}$  is the relative distance at time $t+\delta t$, when the photon is emitted. Where $\delta t$ is the typical decay time for the Rubidium 87 D1 transition.

Transforming to the center of mass and relative velocity coordinates the velocity distribution is a Maxwell Boltzmann distribution of particles with a reduced mass $\mu=m/2$, a kinetic energy of $E_{k}=\frac{\mean{{\vec{{p_{r}}}}^{2}}}{2\mu}$ , momentum $\vec{p_{r}}=\mu\cdot\vec v$ , and relative velocity $\vec v$. 
The velocity distribution for the relative particle, 
\begin{equation}
f\brac v=\sqrt{\f{\mu}{2\pi k_{B}T}}e^{-\f{\mu v^{2}}{2k_{B}T}} 
\end{equation}
The initial relative distance is assumed to be the mean distance for a density $\rho\brac x$ , where $x$  is the spatial position in the trap.
\begin{equation}
\label{eq: mean distance}
r_{i}=\brac{\rho\brac x}^{-1/3} 
\end{equation}
The final relative atomic distance, $r_{f}$ , can be written in terms of the relative velocity $v$ ; $r_{f}=r_{i}+v\cdot\delta t$. Making an ansatz of Eq. \ref{eq: mean distance}.
\begin{equation}
\Delta E=C_{3}\left({\rho\brac x-\f 1{\brac{r_{i}+v\cdot\delta t}^{3}}}\right)+C_{6}\left({\f 1{\brac{r_{i}+v\cdot\delta t}^{6}}-\rho^{2}\brac x}\right)
\end{equation}
Since $r_{i}\gg v\cdot\delta t$  (in the density range discussed) we can expand in a Taylor series up to the first term. 

\begin{equation}
\f 1{\brac{r_{i}+v\cdot\delta t}^{n}}\approx\f 1{r_{i}^{n}}\cdot\left({1-n\f{v\cdot\delta t}{r_{i}}}\right) 
\end{equation}
The energy gap is reduced to;

\begin{equation}
\Delta E=3\brac{\rho\brac x}^{4/3}v\cdot\delta t\brac{C_{3}-2\rho\brac xC_{6}}=C\cdot v 
\end{equation}

\begin{equation}
C=3\brac{\rho\brac x}^{4/3}\delta t\brac{C_{3}-2\rho\brac xC_{6}} 
\end{equation}
In the first order approximation the energy change and the relative velocity are linearly dependent.
The distribution function in velocity translates to an energy distribution function, for $E=\hbar\brac{\omega_{f}-\omega_{i}}$.
\begin{equation}
f\brac{E=\hbar\omega_{f}}=N_{norm}e^{-\f{\mu\hbar^{2}\brac{\omega_{f}-\omega_{i}}^{2}}{2C^{2}k_{B}T}}
\end{equation}

\begin{equation}
N_{norm}=\sqrt{\f 1{\pi}\f{\mu}{2C^{2}k_{B}T}}
\end{equation}
The variance of the function $f\brac{\hbar\omega_{f}}$  can be used to calculate the light phase space diffusion variable of energy transfer, $\mathcal{D}_{E}\brac{\rho_{par},T_{par}}=\f{var\brac{f\brac E}}{\delta t}$ , arising from the interaction of the particle and photons, including only photons which are absorbed.

\begin{equation}
Var\brac{f\brac E}=\f{C^{2}k_{B}T}{\mu} 
\end{equation}

\begin{equation}
C=3\brac{\rho\brac x}^{4/3}\delta t\brac{C_{3}-2\rho\brac xC_{6}} 
\end{equation}

\begin{equation}
\mathcal{D}_{E}\brac{\rho,T}=\f{\sbrac{3\brac{\rho\brac x}^{4/3}\brac{C_{3}-2\rho\brac xC_{6}}}^{2}\delta t\cdot k_{B}T}{\mu} 
\end{equation}
For the diffusion in momentum using the photon energy relation, $E=p\cdot c$ , the diffusion variable in momentum is given by: $\mathcal{D}=\f 1{c^{2}}\mathcal{D}_{E}$ 

\subsection{Random phase approach for calculating the absorption probability function}
We use the Random phase approach as an efficient scheme for propagating a thermal state $\hat{\rho}$.  A thermal state is an incoherent state which undergoes coherent time evolution. In this case a direct approach is a full solution of the Liouville von Neumann equation in the Schr\"odinger picture,

\begin{equation}
i\hbar\pd{\rho}t=\sbrac{H,\rho} 
\end{equation}
For a time evolution operator $\hat{U}\brac t=e^{-\f i{\hbar}\hat{H}t}$  the dynamics can be captured by the equation $\rho\brac t=\hat{U}\brac{t,0}\rho\brac 0\hat{U}^{\dagger}\brac{t,0}$. When a wide range of energy states are populated the direct solution of the initial state can be difficult and time consuming. An alternative approach decomposes the initial thermal state to random phase Gaussian wave functions. Time evolution can be calculated on each realization and averaged to assemble the thermal state at time $t$. A detailed description follows.
For a high number of realizations the random phases cancel each other leaving no effect on the the desired calculation. This is the underlying principle of the method. For a general random phase $e^{i\theta_{\alpha}}$  where $N\gg1$  we can write the Cronicer delta function as:
\begin{equation}
\f 1N\sum_{k=1}^{N}e^{i\brac{\theta_{\alpha}^{k}-\theta_{\beta}^{k}}}=\delta_{\alpha\beta} 
\end{equation}
$k$  labels a set random angle, each angle given for each basis state , $\alpha$  and $\beta$. If $\alpha=\beta$, $k_{\alpha}=k_{\beta}$  for all $k$ , we get the unity, for any other case the equality converges to zero as $\f 1{\sqrt{N}}$.
This characteristic allows a composition of the operator with an arbitrary complete orthonormal basis $\{\ket{\alpha}\}$  and the random phases $\{e^{i\theta_{\alpha}^{k}}\}$  . We define a thermal random wave function $\ket{\psi_{\alpha}^{k}}=e^{i\theta^{k}_{\alpha}}\ket{\alpha}$  and an accumulated wave function $\ket{\Psi^{k}}=\sum_{\alpha}\ket{\psi_{\alpha}^{k}}=\sum_{\alpha}e^{i\theta^{k}_{\alpha}}\ket{\alpha}$ 

\begin{equation}
\hat{1}=\f 1N\sum_{k=1}^{N}\ket{\Psi^{k}}\bra{\Psi^{k}}=\sum_{\alpha,\beta}\ket{\alpha}\bra{\beta}\f 1N\sum_{k=1}^{N}e^{i\brac{\theta_{\alpha}^{k}-\theta_{\beta}^{k}}}=\f 1N\sum_{k=1}^{N}\sum_{\alpha,\beta}\ket{\psi_{\alpha}^{k}}\bra{\psi_{\beta}^{k}} 
\end{equation}
Therefore the thermal state at time $t=0$  is;
\begin{eqnarray*}
\label{eq: rho random phase}
\hat{\rho}=\hat{\rho}\cdot\hat{1}=\f 1Ze^{-\f{\hat{H}\beta}2}e^{-\f{\hat{H}\beta}2}\sum_{\alpha,\beta}\ket{\alpha}\bra{\beta}\f 1N\sum_{k=1}^{N}e^{i\brac{\theta_{\alpha}-\theta_{\beta}}}  \\
=\f 1Z\f 1N\sum_{k=1}^{N}\sum_{\alpha,\beta}e^{i\theta_{\alpha}^{k}}e^{-\f{E_{\alpha}\beta}2}\ket{\alpha}\bra{\beta}e^{-\f{E_{\beta}\beta}2}e^{-i\theta_{\beta}^{k}}=\f 1Z\f 1N\sum_{k=1}^{N}\ket{\varphi^{k}}\bra{\varphi^{k}} 
\end{eqnarray*}
while the thermal random wave functions are $\ket{\varphi^{k}}=\sum_{\alpha}e^{-\f{E_{\alpha}\beta}2+i\theta_{\alpha}^{k}}\ket{\alpha}$ , and the temperature dependence given by $\beta=\f 1{k_{B}T}$  .

The thermally averaged time dependent states , $\hat{\rho}\brac t$ , can be calculated by the same process, decomposed to time dependent thermal random wave functions $\ket{\varphi^{k}\brac t}$. We obtain the thermal state $\hat{\rho}\brac t$  by propagating $N$  accumulated thermal random function, $\ket{\varphi^{k}\brac t}=\hat{U}\brac{t,0}\ket{\varphi^{k}}$ , and taking an average defined by equation \ref{eq: rho random phase}.
Taking a closer look at a single thermal random wave function $\ket{\varphi_{\alpha}^{k}}=e^{-\f{E_{\alpha}\beta}2+i\theta_{\alpha}^{k}}\ket{\alpha}$,  
$\{\ket{\alpha}\}$  are chosen to be the momentum state basis. For small potential energy the state can be written as;
\begin{equation}
e^{-\f{\hat{H}\beta}2+i\theta_{p}^{k}}\ket p\approx e^{-\f{p^{2}\beta}{2m}+i\theta_{p}^{k}}\ket p=e^{-\f{p^{2}}{2mk_{B}T}+ipR_{0}}\ket p 
\end{equation}
having defined $\theta_{p}^{k}=pR_{0}$  in the second equalization.

The thermal random state, for the range of high kinetic energy or weak interactions, is a thermal Gaussian with a variance $mk_{B}T$  and an additional random phase. In the position representation the wave function has a form of a Gaussian displaced by $R_{0}$, $\ket{\varphi_{r}^{k}}=e^{-\frac{1}{2}mk_{B}T\brac{r-R_{0}}^{2}}$
In the position representation the random phase approach leads to a decomposition of the initial thermal state to many thermal Gaussian wave functions centered randomly in space. The validity of such an approximation for a two-body interaction holds only for large $r$  where the potential is weak. 

\break

%\bibliography{citations}

%\bibliographystyle{unsrt}

%\printbibliography

\end{document}